\documentclass{article}

\usepackage{amssymb,amsfonts,amsthm}
\usepackage{graphicx}
\usepackage{subfigure}
\usepackage{amssymb}
\usepackage{amsthm}
\usepackage{amsmath}
\usepackage{bm}             
\usepackage[mathscr]{eucal}

\usepackage{cancel}

\usepackage{psfrag}

\usepackage{wasysym}

\renewcommand{\u}{\mathsf{u}}

\newcommand{\BI}{\bm{B}_{\,\mathrm{I}}}
\newcommand{\BII}{\bm{B}_{\mathrm{II}}}
\newcommand{\BVI}{\bm{B}_{\mathrm{VI}_0}}
\newcommand{\BVII}{\bm{B}_{\mathrm{VII}_0}}
\newcommand{\BVIII}{\bm{B}_{\mathrm{VIII}}}
\newcommand{\BIX}{\bm{B}_{\mathrm{IX}}}

\newcommand{\overlineBVII}{\overline{\bm{B}}_{\mathrm{VII}_0}}
\newcommand{\overlineBIX}{\overline{\bm{B}}_{\mathrm{IX}}}

\newcommand{\checkP}{\check{\mathrm{P}}}

\newcommand{\weg}{\:\,}

\newcommand{\bom}{\bm{\omega}}

\newcommand{\diag}{\mathop{\mathrm{diag}}}

\newcommand{\spur}{\mathop{\mathrm{tr}}}
\newcommand{\textfrac}[2]{{\textstyle \frac{#1}{#2}}}

\newcommand{\norDelta}{\delta}
\newcommand{\barDelta}{\Delta}

\newcommand{\lin}{l_{\mathrm{in}}}
\newcommand{\lout}{l_{\mathrm{out}}}

\theoremstyle{plain}
\newtheorem{theorem}{Theorem}[section]

\newtheorem*{conjecture}{Conjecture}
\newtheorem*{definition}{Definition}

\theoremstyle{remark}

\def\pmb#1{\setbox0=\hbox{$#1$}%
  \kern-.025em\copy0\kern-\wd0
  \kern.05em\copy0\kern-\wd0
  \kern-.025em\raise.0433em\box0}
\def\pmbs#1{\setbox0=\hbox{$\scriptstyle #1$}%
  \kern-.0175em\copy0\kern-\wd0
  \kern.035em\copy0\kern-\wd0
  \kern-.0175em\raise.0303em\box0}

\def\cg{{\cal G}}
\def\nt{\tilde N}
\newcommand{\sfrac}[2]{{\textstyle{\frac{#1}{#2}}}}

\begin{document}


\title{\bf Mixmaster: Fact and Belief}

\author{ \\
{\Large\sc J.\ Mark Heinzle}\thanks{Electronic address:
{\tt mark.heinzle@univie.ac.at}} \\[1ex]
Gravitational Physics, Faculty of Physics, \\
University of Vienna, A-1090 Vienna, Austria \\
and \\
Mittag-Leffler Institute of the Royal Swedish Academy of Sciences\\
S-18260 Djursholm, Sweden
\and \\
{\Large\sc Claes Uggla}\thanks{Electronic address:
{\tt claes.uggla@kau.se}} \\[1ex]
Department of Physics, \\
University of Karlstad, S-651 88 Karlstad, Sweden \\[2ex] }

\date{}
\maketitle

\begin{abstract}

We consider the dynamics towards the initial singularity of
Bianchi type~IX vacuum and orthogonal perfect fluid models with
a linear equation of state. Surprisingly few facts are known
about the `Mixmaster' dynamics of these models, while at the
same time most of the commonly held beliefs are rather vague.
In this paper, 
we use Mixmaster facts as a base to build an infrastructure
that makes it possible to sharpen the main Mixmaster beliefs.
We formulate explicit conjectures concerning (i) the past
asymptotic states of type~IX solutions and (ii) the relevance
of the Mixmaster/Kasner map for generic past asymptotic
dynamics. The evidence for the conjectures is based on a study
of the stochastic properties of this map in conjunction with
dynamical systems techniques. We use a dynamical systems
formulation, since this approach has so far been the only
successful path to obtain theorems, but we also make
comparisons with the `metric' and Hamiltonian `billiard'
approaches.

\end{abstract}

\newpage

\section{Introduction}\label{intro}

Today, Bianchi type~IX enjoys an almost mythical status in
general relativity and cosmology, which is due to two commonly
held beliefs: (i) Type~IX dynamics is believed to be
essentially understood; (ii) Bianchi type~IX is believed to be
a role model that captures the generic features of generic
spacelike singularities. However, we will illustrate in this
paper that there are reasons to question these beliefs.

The idea that type~IX is essentially understood is a
misconception. In actuality, surprisingly little is known,
i.e., proved, about type~IX asymptotic dynamics; at the same
time there exist widely held, but rather vague, beliefs about
Mixmaster dynamics, oscillations, and chaos, which are
frequently mistaken to be facts. There is thus a need for
clarification: What are the known facts and what is merely
believed about type~IX asymptotics? We will address this issue
in two ways: On the one hand, we will discuss the main rigorous
results on Mixmaster dynamics, the `Bianchi type~IX attractor
theorem', and its consequences; in particular, we will point
out the limitations of these results. On the other hand, we
will provide the infrastructure that makes it possible to
sharpen commonly held beliefs; based on this framework we will
formulate explicit refutable conjectures.

Historically, Bianchi type IX vacuum and orthogonal perfect fluid
models entered the scene in the late sixties through the work of
Belinskii, Khalatnikov and Lifshitz%
\footnote{We will refer
to the authors and their work as BKL.}
\cite{lk63,bkl70} and Misner and
Chitr\'e~\cite{mis69a,mis69b,grav73,chi72}. BKL attempted to
understand the detailed nature of singularities and were led to
the type IX models via a rather convoluted route, while Misner
was interested in mechanisms that could explain why the
Universe today is almost isotropic. BKL and Misner
independently, by means of quite different methods, reached the
conclusion that the temporal behavior of the type IX models
towards the initial singularity can be described by sequences
of anisotropic Kasner states, i.e., Bianchi type I vacuum
solutions. These sequences are determined by a discrete map
that leads to an oscillatory anisotropic behavior, which
motivated Misner to refer to the type IX models as Mixmaster
models~\cite{mis69a,mis69b}. This discrete map, the Kasner map,
was later shown to be associated with stochasticity and
chaos~\cite{khaetal85,bar82,chebar83}, a property that has
generated considerable interest---and confusion, see,
e.g.,~\cite{waiell97,col03,aizetal97,ber02,elshen87,cheetal05,hobetal94,corlev97,corlev97b,rugh90,motlet01}
and references therein. A sobering thought: All claims about
chaos in Einstein's equations rest on the (plausible) belief
that the Kasner map actually describes the asymptotic dynamics
of Einstein's equations; as will be discussed below, this is
far from evident (despite being plausible) and has not been
proved so far.

More than a decade after BKL's and Misner's investigations a
new development took place: Einstein's field equations in the
spatially homogeneous (SH) case were reformulated in a manner
that allowed one to apply powerful dynamical systems
techniques~\cite{col71,bognov73,bog85}; gradually a picture of
a hierarchy of invariant subsets emerged where monotone
functions restricted the asymptotic dynamics to boundaries of
boundaries, see~\cite{waiell97} and references therein. Based
on work reviewed and developed in~\cite{waiell97} and by
Rendall~\cite{ren97}, Ringstr\"om eventually produced the first
major proofs about asymptotic type~IX
dynamics~\cite{rin00,rin01}. This achievement is remarkable,
but it does not follow that all questions are settled. On the
contrary, so far nothing is rigorously known, e.g., about
dynamical chaotic properties (although there are good grounds
for beliefs), nor has the role of type~IX models in the context
of generic singularities been
established~\cite{heiuggproof,uggetal03,andetal05,heietal07}.

The outline of the paper is as follows. In Section~\ref{basic}
we briefly describe the Hubble-normalized dynamical systems
approach and establish the connection with the metric approach.
For simplicity we restrict ourselves to the vacuum case and the
so-called orthogonal perfect fluid case, i.e., the fluid flow
is orthogonal w.r.t.\ the SH symmetry surfaces; furthermore, we
assume a linear equation of state. In Section~\ref{subsets} we
discuss the levels of the Bianchi type~IX so-called Lie
contraction hierarchy of subsets, where we focus on the Bianchi
type~I and type~II subsets. In Section~\ref{nongeneric} we
present the results of the local analysis of the fixed points
of the dynamical system and discuss the stable and unstable
subsets of these points which are associated with non-generic
asymptotically self-similar behavior. Section~\ref{maps} is
devoted to a study of the network of sequences of heteroclinic
orbits (heteroclinic chains) that is induced by the dynamical
system on the closure of the Bianchi type~II vacuum boundary of
the type~IX state space (which we refer to as the Mixmaster
attractor subset). These sequences of orbits are associated
with the Mixmaster map, which in turn induces the Kasner map
and thus the Kasner sequences. We analyze the properties of
non-generic Kasner sequences and discuss the stochastic
properties of generic sequences. In Section~\ref{furthermix} we
discuss the main `Mixmaster facts': Ringstr\"om's `Bianchi
type~IX attractor theorem'~\cite{rin00,rin01},
Theorem~\ref{rinthm}, and a number of consequences that follow
from Theorem~\ref{rinthm} and from the results on the
Mixmaster/Kasner map. In addition, we introduce and discuss the
concept of `finite Mixmaster shadowing'. In the subsection
`Attractor beliefs' of Section~\ref{stochasticbeliefs} we
formulate two conjectures that reflect commonly held beliefs
about type~IX asymptotic dynamics and list some open issues
that are directly connected with these conjectures. In the
subsection `Stochastic beliefs' we address the open question of
which role the Mixmaster/Kasner map and its stochastic
properties actually play in type~IX asymptotic dynamics. This
culminates in the formulation, and discussion, of two
`stochastic' conjectures. In Section~\ref{billiard} we present
the Hamiltonian billiard formulation, see~\cite{grav73,chi72}
or~\cite{dametal03}; we demonstrate that this approach yields a
`dual' formulation of the asymptotic dynamics. We point out
that the billiard approach is a formidable heuristic picture,
but fails to turn beliefs into facts. We conclude in
Section~\ref{concl} with a
discussion of the main themes of this paper. 
Throughout this paper we use units so that $c=1$ and $8\pi G=1$,
where $c$ is the speed of light and $G$ the gravitational constant.

\section{Basic equations}
\label{basic}

We consider vacuum or orthogonal perfect fluid SH Bianchi
type~IX models (i.e., the fluid 4-velocity is assumed to be
orthogonal to the SH symmetry surfaces) with a
linear equation of state; we require
the energy conditions (weak/strong/dominant) to hold, i.e.,
$\rho > 0$ and
\begin{equation}\label{wassum}
-\sfrac{1}{3} < w < 1 \:,
\end{equation}
where  $w = p/\rho$, and where $\rho$ and $p$ are the energy
density and pressure of the fluid, respectively.
By~\eqref{wassum} we exclude the special cases
$w=-\textfrac{1}{3}$ and
$w=1$, where the energy conditions are only marginally satisfied.%
\footnote{Note that the well-posedness of
  the Einstein equations (for solutions without symmetry)
  has been questioned in the case $-1/3<w<0$, see~\cite{friren00}. The case $w=1$ is
  known as the stiff fluid case, for which the speed of sound is equal to
  the speed of light. The asymptotic dynamics of stiff fluid solutions is
  simpler than the oscillatory behavior characterizing the models with
  range $-\sfrac{1}{3} < w < 1$, and well understood~\cite{rin01,andren01}.
  (In the terminology introduced below,
  the stiff fluid models are asymptotically self-similar.)
  We will therefore refrain from discussing the stiff fluid case in this paper.}%
%
%

As is well known, see, e.g.,~\cite{waiell97,rin01} and
references therein, for these models there exists a
symmetry-adapted \mbox{(co-)}frame
$\{\hat{\bom}^1,\hat{\bom}^2,\hat{\bom}^3\}$,
\begin{subequations}\label{bixform}
\begin{gather}
\label{structconst}
d\hat{\bom}^1  =  -\hat{n}_1 \, \hat{\bom}^2\wedge \hat{\bom}^3\:,\quad
d\hat{\bom}^2  =  -\hat{n}_2 \, \hat{\bom}^3\wedge \hat{\bom}^1\:,\quad
d\hat{\bom}^3  =  -\hat{n}_3 \, \hat{\bom}^1\wedge \hat{\bom}^2\:,
\intertext{with $\hat{n}_1 =1$, $\hat{n}_2 = 1$, $\hat{n}_3 =1$, such that the type~IX metric
takes the form}
\label{threemetric}  {}^4\mathbf{g}  = -dt\otimes dt +
g_{11}(t)\:\hat{\bom}^1\otimes \hat{\bom}^1 +
g_{22}(t)\:\hat{\bom}^2\otimes \hat{\bom}^2 +
g_{33}(t)\:\hat{\bom}^3\otimes \hat{\bom}^3\:.
\end{gather}
\end{subequations}
Hence, the type~IX models naturally belong to the so-called
class~A Bianchi models, 
see Table~\ref{classAmodels}.
Let
\begin{equation}
\label{nsandmetric}
n_1(t) := \hat{n}_1 \,\frac{g_{11}}{\sqrt{\det g}} \:, \quad
n_2(t) := \hat{n}_2 \,\frac{g_{22}}{\sqrt{\det g}} \:,\quad
n_3(t) := \hat{n}_3 \,\frac{g_{33}}{\sqrt{\det g}} \:,
\end{equation}
where $\det g = g_{11} g_{22} g_{33}$. Furthermore, define
\begin{equation}
\theta = -\spur k\qquad\text{and}\qquad \sigma^\alpha_{\weg \beta} =
-k^{\alpha}_{\weg \beta} + \sfrac{1}{3} \spur k \:\delta^\alpha_{\weg \beta}=
\diag(\sigma_1,\sigma_2,\sigma_3) \quad \left(\Rightarrow
\sum\nolimits_\alpha \sigma_\alpha = 0\right)\:,
\end{equation}
where $k_{\alpha\beta}$ denotes the second fundamental form
associated with~\eqref{bixform} of the SH hypersurfaces
$t=\mathrm{const}$. The quantities $\theta$ and
$\sigma_{\alpha\beta}$ can be interpreted as the expansion and
the shear, respectively, of the normal congruence of the SH
hypersurfaces. In a cosmological context it is customary to
replace $\theta$ by the Hubble variable $H=\theta/3 = -\spur
k/3$; this variable is related to changes of the spatial volume
density according to $d\sqrt{\det g}/d t = 3 H \sqrt{\det g}$.
Evidently, in Bianchi type~IX (and type VIII) there is a
one-to-one correspondence between the `orthonormal frame
variables' $(H, \sigma_\alpha, n_\alpha)$ (with $\sum_\alpha
\sigma_\alpha =0$) and $(g_{\alpha\beta}, k_{\alpha\beta})$; in
particular, the metric $g_{\alpha\beta}$ is obtained from
$(n_1,n_2,n_3)$ via~\eqref{nsandmetric}. (For the lower Bianchi
types~I--$\mathrm{VII}_0$, some of the variables
$(n_1,n_2,n_3)$ are zero, cf.~\eqref{nsandmetric}; in this
case, the other frame variables, i.e., $(H, \sigma_\alpha)$,
are needed as well to reconstruct the metric;
see~\cite{janugg99} for a group theoretical approach.)

\begin{table}
\begin{center}
\begin{tabular}{|c|ccc|}
\hline Bianchi type & $\hat{n}_\alpha$ &  $\hat{n}_\beta$ & $\hat{n}_\gamma$ \\ \hline
I & $0$ & $0$ & $0$ \\
II & $0$ & $0$ & $+$ \\
$\mathrm{VI}_0$ & $0$ & $-$ & $+$ \\
$\mathrm{VII}_0$ & $0$& $+$ & $+$ \\
$\mathrm{VIII}$ & $-$& $+$& $+$ \\
$\mathrm{IX}$ & $+$& $+$& $+$ \\\hline
\end{tabular}
\caption{The class~A Bianchi types are characterized by
different signs of the structure constants $(\hat{n}_\alpha, \hat{n}_\beta,
\hat{n}_\gamma)$, where $(\alpha\beta\gamma)$ is any permutation of
$(123)$. In addition to the above representations there exist
equivalent representations associated with an overall change of
sign of the structure constants; e.g., another type IX representation is
$(---)$.} \label{classAmodels}
\end{center}
\end{table}
%

In the Hubble-normalized
dynamical systems approach
we define dimensionless
orthonormal frame variables according to
\begin{equation}\label{Hnormvars}
(\Sigma_\alpha, N_\alpha) = (\sigma_\alpha, n_\alpha)/H\:, \qquad
\Omega = \rho/(3H^2)\:.
\end{equation}
In addition we introduce a new dimensionless time variable
$\tau$ according to $d\tau/dt=H$. Like the cosmological time
$t$, the time $\tau$ is directed towards the future; however,
to make contact with the well established convention that uses
a past-directed `time' for the discrete Mixmaster map, see
Section~\ref{maps}, it will occasionally become necessary to
use an inverse time $\tau_- = -\tau$ instead of $\tau$ itself.

For all class A models except type~IX the Gauss constraint
guarantees that $H$ remains positive if it is positive
initially. In Bianchi type~IX, however, it is known from a
theorem by Lin and Wald~\cite{linwal89} that all type~IX vacuum
and orthogonal perfect fluid models with $w\geq 0$ first expand
($H>0$), reach a point of maximum expansion ($H=0$), and then
recollapse ($H<0$).%
\footnote{In the locally rotationally symmetric case it has been
  proved that the range of $w$ can be
  extended to $w>-\frac{1}{3}$, see~\cite{heietal05}. There are
  good reasons to believe that the assumption of local rotational symmetry
  is superfluous, but this has not been established yet.}
The variable transformation~\eqref{Hnormvars} breaks down at
the point of maximum expansion in the type~IX case; however,
the variables $(\Sigma_\alpha, N_\alpha)$ correctly describe
the dynamics in the expanding phase, which we will focus on
henceforth.

When the Einstein field equations are reformulated in terms of
$H$ and the Hubble-normalized variables $(\Sigma_\alpha,
N_\alpha)$ it follows for dimensional reasons that the
equation for the single variable with dimension, $H$,
\begin{equation}\label{H}
H^\prime = -(1+q)H\:,
\end{equation}
decouples from the remaining dimensionless
equations~\cite{waiell97}; here and henceforth a prime denotes
the derivative $d/d\tau$. The equations for $(\Sigma_\alpha,
N_\alpha)$ form the following coupled system~\cite{waiell97}:
\begin{subequations}\label{IXeq}
\begin{align}
\label{sig}
\Sigma_\alpha^\prime & =  -(2-q)\Sigma_\alpha - {}^{3}\!S_\alpha\:, \\[0.5ex]
\label{n}
N_\alpha^\prime & =  (q+2\Sigma_\alpha)\,N_\alpha
\qquad\qquad\qquad \text{(no sum over $\alpha$)}\:,
\end{align}
\end{subequations}
where
\begin{subequations}
\begin{alignat}{2}
\label{q}
q & = 2\Sigma^2 + \sfrac{1}{2}(1+3w)\Omega \:,
&  & \Sigma^2 = \sfrac{1}{6}(\Sigma_1^2 + \Sigma_2^2 + \Sigma_3^2)\:,\\
\label{threecurv}
\text{and}\quad  {}^{3}\!S_\alpha  & = \sfrac{1}{3}\left[ N_\alpha(2N_\alpha
- N_\beta - N_\gamma) - (N_\beta - N_\gamma)^2 \right],  & \quad &
(\alpha\beta\gamma) \in \left\{(123),(231),(312)\right \}\:.
\end{alignat}
\end{subequations}
Note that the `deceleration parameter' $q$ is non-negative
because of the assumption $w>-1/3$.

Apart from the trivial constraint
$\Sigma_1 + \Sigma_2 + \Sigma_3 = 0$, there exists the Gauss
constraint
\begin{equation}
\label{gauss}
\Sigma^2 +
\sfrac{1}{12} \Big[ N_1^2 + N_2^2 + N_ 3^2 -
2 \underbrace{\left( N_1N_2 + N_2 N_3 + N_3N_1 \right)}_{\text{\normalsize
$\Delta_{\mathrm{II}}$}}
\Big] + \Omega = 1 \:,
\end{equation}
which is used to globally solve for $\Omega$ when $\Omega\neq
0$. Accordingly, the reduced state space is given as the space
of all $(\Sigma_1,\Sigma_2,\Sigma_3)$ and $(N_1,N_2,N_3)$ such
that $\Sigma_1 + \Sigma_2 + \Sigma_3 = 0$ and
\begin{equation}\label{statesp}
\Sigma^2 + \sfrac{1}{12} \Big[ N_1^2 + N_2^2 + N_ 3^2 -
2
\left( N_1N_2 + N_2 N_3 + N_3N_1 \right)
\Big] \leq 1 \:,
\tag{\ref{gauss}${}'$}
\end{equation}
which follows from~\eqref{gauss} under the assumption that
$\Omega \geq 0$ ($\rho\geq 0$).
It follows that the dimensionless state space of
the Bianchi type~IX orthogonal perfect fluid models with a
linear equation of state
is 5-dimensional,%
\footnote{It is common to globally solve $\Sigma_1 + \Sigma_2 + \Sigma_3 = 0$
  by introducing new variables according to
  $\Sigma_1 = -2 \Sigma_+$, $\Sigma_2 =
  \Sigma_+ - \sqrt{3} \Sigma_-$, $\Sigma_3 = \Sigma_+ + \sqrt{3} \Sigma_-$,
  which yields $\Sigma^2 = \Sigma_+^2 + \Sigma_-^2$. However, since this breaks the
  permutation symmetry of the three spatial axes (exhibited by type~IX models),
  we choose to retain the variables $\Sigma_1$, $\Sigma_2$, $\Sigma_3$.}
while the state space of the vacuum models (i.e., $\Omega=0$)
is 4-dimensional. The same is true for Bianchi type~VIII, while
the state spaces of the remaining class~A Bianchi models have
less degrees of freedom; see Table~\ref{Bianchistatespaces}.
Once the dynamics in the dimensionless state space is
understood, $H$ is obtained from a quadrature by
integrating~\eqref{H}, which allows one to reconstruct the
metric.

\begin{table}
\begin{center}
\begin{tabular}{|ccc|c|c|}
\hline Bianchi type & Symbol & Range of $(N_\alpha, N_\beta,
N_\gamma)$ & State space & $\mathsf{D}$ \\ \hline I & $\BI$ &
$N_\alpha =0$, $N_\beta = 0$,
$N_\gamma = 0$ & $\Sigma^2 \leq 1$ & 2 \\
II & $\BII$ &  $N_\alpha =0$, $N_\beta = 0$, $N_\gamma > 0$
& $\Sigma^2 +\sfrac{1}{12} N_\gamma^2 \leq 1$ & 3\\
$\mathrm{VI}_0$ & $\BVI$ & $N_\alpha = 0$, $N_\beta <0$, $N_\gamma > 0$
& $\Sigma^2 +\sfrac{1}{12} [N_\beta -  N_\gamma]^2 \leq 1$ & 4 \\
$\mathrm{VII}_0$ & $\BVII$ &  $N_\alpha = 0$, $N_\beta > 0$, $N_\gamma > 0$
& $\Sigma^2 +\sfrac{1}{12} [N_\beta - N_\gamma]^2 \leq 1$ & 4\\
$\mathrm{VIII}$ & $\BVIII$ & $N_\alpha < 0$, $N_\beta > 0$, $N_\gamma > 0$
& \parbox{3.8cm}{Eq.~\eqref{statesp} ($\Rightarrow \Sigma^2 < 1$)}& 5 \\
$\mathrm{IX}$ & $\BIX$ & $N_\alpha > 0$, $N_\beta > 0$, $N_\gamma
> 0$ & \parbox{3.8cm}{Eq.~\eqref{statesp} ($\Rightarrow \Sigma^2 \leq
1 + \Delta$)}& 5 \\\hline
\end{tabular}
\caption{The dimensionless state spaces associated with class~A
Bianchi models; here, $(\alpha\beta\gamma)$ is any permutation
of $(123)$. In addition to the above representations there
exist equivalent representations associated with an overall
change of sign of the variables $(N_1,N_2,N_3)$. The quantity
$\mathsf{D}$ denotes the dimension of the state space (in the
fluid case); the dimensionality of the state space in the
vacuum cases is given by $\mathsf{D} - 1$.}
\label{Bianchistatespaces}
\end{center}
\end{table}

For all class~A models except type IX the
constraint~\eqref{statesp} implies that $\Sigma^2 \leq 1$; in
Bianchi type~IX, however, $\Sigma^2 > 1$ is possible. For type
IX, define
\begin{equation}\label{Deltadef}
\Delta = \sfrac{1}{4}  (N_1N_2N_3)^{2/3}\:.
\end{equation}
Employing~\eqref{q} and~\eqref{gauss} and using that
\begin{equation}
\sfrac{1}{12} \Big[ N_1^2 + N_2^2 + N_ 3^2 - 2 \Delta_{\mathrm{II}}\Big]
+ \Delta \geq 0\:,
\end{equation}
where equality holds iff $N_1=N_2=N_3$, we find
\begin{equation}\label{ineq}
\Sigma^2 \leq 1 + \Delta\:, \qquad \Omega \leq 1 + \Delta\:, \qquad
2 -q \geq \sfrac{3}{2}(1-w)\Omega - 2 \Delta\:.
\end{equation}
The function $\Delta$ is strictly monotonically increasing along
orbits of Bianchi type~IX. To see this we use~\eqref{IXeq} and compute
\begin{equation}\label{Deltamon}
\Delta^\prime = 2q\Delta \:,\qquad\Delta^{\prime\prime} \Big|_{q = 0} = 0\:,
\qquad
\Delta^{\prime\prime\prime}\Big|_{q = 0} = \sfrac{4}{3}
\left[  {}^3\!S_1^2 + {}^3\!S_2^2 + {}^3\!S_3^2\right] \Delta\:,
\end{equation}
where we note that ${}^3\!S_1^{\,2} + {}^3\!S_2^{\,2} +
{}^3\!S_3^{\,2} > 0$ because of the constraints. In combination
with~\eqref{ineq} it follows that $\Sigma^2$ and $\Omega$ are
\emph{bounded towards the past}.

The right hand side of the reduced system~(\ref{IXeq}) consists
of polynomials of the state space variables and is thus a
regular dynamical system. Solutions of~\eqref{IXeq} of Bianchi
types~I--VIII are global in $\tau$, since~\eqref{gauss} implies
the bounds $\Sigma^2 \leq 1$ and $\Omega \leq 1$ which control
the evolution of $N_\alpha$ in~\eqref{n}. Solutions
of~\eqref{IXeq} of Bianchi type~IX are global towards the past,
since $q + 2 \Sigma_\alpha$ is bounded from below; this follows
from~\eqref{q}, which yields $q\geq 2\Sigma^2$, so that $q +
2\Sigma_\alpha \geq -2 + \frac{1}{2} (\Sigma_\alpha +2)^2 +
\frac{1}{6} (\Sigma_\beta -\Sigma_\gamma)^2$. The decoupled
equation for $H$, cf.~\eqref{H}, yields that $H\rightarrow
\infty$ as $\tau\rightarrow -\infty$, because $q$ is
non-negative. Since $q$ is bounded as $\tau\rightarrow
-\infty$, the asymptotics of $H$ can be bounded by exponential
functions from above and below. It follows that the equation $d
t/d\tau = H^{-1}$ can be integrated to yield $t$ as a function
of $\tau$ such that $t\rightarrow 0$ as $\tau\rightarrow
-\infty$.

In addition to~\eqref{IXeq} it is useful to also consider
an auxiliary equation for the matter quantity $\Omega$,
\begin{equation}
\label{Omegaeq}
\Omega^\prime =  [2q - (1 + 3w)]\Omega\:.
\end{equation}
Making use of~\eqref{Hnormvars} and~\eqref{H} we conclude that for
all orthogonal perfect fluid models with a linear equation of state
we have $\rho \propto \exp\left(-3[1+w]\tau\right)$, and hence
$\rho\rightarrow\infty$ as $\tau\rightarrow -\infty$, which yields a
past singularity. 
The divergence of $\rho$ can also be directly read off from the matter
equation $\nabla_a T^{a b} = 0$.

\section{The Bianchi type IX Lie contraction hierarchy}
\label{subsets}

\begin{figure}[ht]
\psfrag{a}[cc][cc]{$\mathcal{B}_{N_1N_2N_3}$}
\psfrag{b}[cc][cc]{$\mathcal{B}_{N_1N_2}$}
\psfrag{c}[cc][cc]{$\mathcal{B}_{N_2N_3}$}
\psfrag{d}[cc][cc]{$\mathcal{B}_{N_1N_3}$}
\psfrag{e}[cc][cc]{$\mathcal{B}_{N_1}$}
\psfrag{f}[cc][cc]{$\mathcal{B}_{N_2}$}
\psfrag{g}[cc][cc]{$\mathcal{B}_{N_3}$}
\psfrag{h}[cc][cc]{$\mathcal{B}_\emptyset$}
\psfrag{i}[rc][cc]{$\BIX$}
\psfrag{j}[rc][cc]{$\BVII$}
\psfrag{k}[rc][cc]{$\BII$}
\psfrag{l}[rc][cc]{$\BI$} \centering{
  \includegraphics[height=0.45\textwidth]{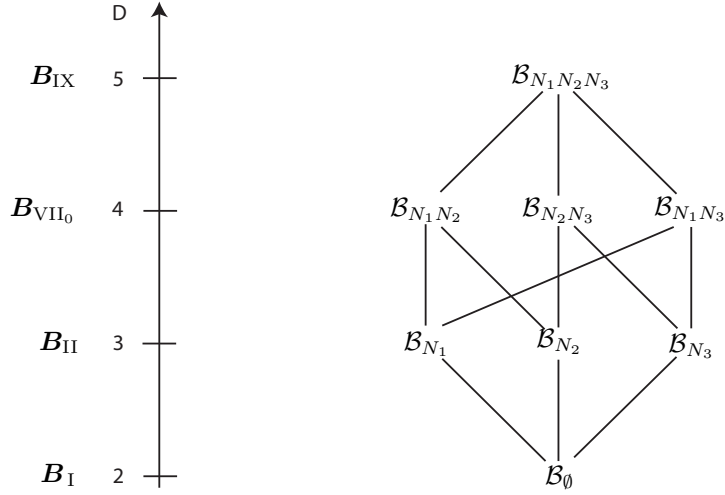}}
\caption{Subset contraction diagram for Bianchi type IX.
$\mathsf{D}$ denotes the dimension of the dimensionless state space
for the various models with an orthogonal perfect fluid with linear
equation of state; the associated vacuum subsets have one dimension
less; see also Table~\ref{Bianchistatespaces}. The notation is such
that the subscript of $\mathcal{B}_{\star}$ denotes the non-zero
variables, e.g. $\mathcal{B}_{N_1}$ denotes the type
$\mathrm{II}$
subset with $N_1>0$, $N_2 =0$ and $N_3 = 0$.}
\label{contraction}
\end{figure}

The Bianchi type~IX state space $\BIX$ is characterized by the
conditions $N_1 > 0$, $N_2 > 0$, $N_3 >0$.
We write $\BIX =
\mathcal{B}_{N_1N_2N_3}$. The notation is such that the subscript
denotes the non-zero variables among $\{N_1, N_2, N_3\}$. Setting
one or more of these variables to zero (which corresponds to Lie
contractions~\cite{jan01}) yields invariant boundary subsets which
describe more special Bianchi types. Since the type IX models
exhibit discrete symmetries associated with axes permutations, the
contractions generate all possible representations of the more
special (Lie contracted) Bianchi types (which are associated with such
permutations): 
The Bianchi type~$\mathrm{VII}_0$ subspace $\BVII$ is given by
the disjoint union of three equivalent sets, $\BVII =
\mathcal{B}_{N_1 N_2} \cup \mathcal{B}_{N_2 N_3} \cup
\mathcal{B}_{N_3 N_1}$, where, e.g., $\mathcal{B}_{N_1N_2}$
denotes the type~$\mathrm{VII}_0$ subset with $N_1>0$, $N_2 >0$
and $N_3 = 0$; the Bianchi type~$\mathrm{II}$ subspace $\BII$
by the union $\BII = \mathcal{B}_{N_1} \cup \mathcal{B}_{N_2}
\cup \mathcal{B}_{N_3}$; the Bianchi type~$\mathrm{I}$ subspace
$\BI$ by $\BI = \mathcal{B}_{\emptyset}$. Note that the Bianchi
type~$\mathrm{VI}_0$ subspace does not appear as a boundary
subset of $\BIX$.
A Bianchi subset contraction diagram for
type IX is given in Figure~\ref{contraction}.

Each set of the Lie contraction hierarchy is the union of
an invariant vacuum subset, i.e., $\Omega = 0$, and
an invariant fluid subset, i.e., $\Omega > 0$.
To refer to a vacuum [fluid] subset of a Bianchi set $\mathcal{B}_\star$
we use the notation $\mathcal{B}_\star^{\mathrm{vac.}}$ [$\mathcal{B}_\star^{\mathrm{fl.}}$].
In this spirit, e.g., the type~II subset decomposes as
$\mathcal{B}_{N_1} = \mathcal{B}_{N_1}^{\mathrm{vac.}} \cup \mathcal{B}_{N_1}^{\mathrm{fl.}}$.

In the following we analyze the boundary subsets to the extent
needed in order to understand the asymptotic type~IX dynamics.

\subsection*{The Bianchi type I subset}

\begin{figure}[ht]
\psfrag{a}[cc][cc]{$\Sigma_1$} \psfrag{b}[cc][cc]{$\Sigma_2$}
\psfrag{c}[cc][cc]{$\Sigma_3$} \psfrag{d}[cc][cc]{$(123)$}
\psfrag{e}[cc][cc]{$(213)$} \psfrag{f}[cc][cc]{$(231)$}
\psfrag{g}[cc][cc]{$(321)$} \psfrag{h}[cc][cc]{$(312)$}
\psfrag{i}[cc][cc]{$(132)$} \psfrag{k}[cc][cc]{$\mathrm{T}_1$}
\psfrag{l}[cc][cc]{$\mathrm{T}_2$}
\psfrag{m}[cc][cc]{$\mathrm{T}_3$}
\psfrag{n}[cc][cc]{$\mathrm{Q}_1$}
\psfrag{o}[cc][cc]{$\mathrm{Q}_2$}
\psfrag{p}[cc][cc]{$\mathrm{Q}_3$}
\psfrag{s}[cc][cc]{$\mathrm{M}_1$}
\psfrag{t}[cc][cc]{$\mathrm{M}_2$}
\psfrag{u}[cc][cc]{$\mathrm{M}_3$}
\psfrag{x}[cc][cc]{$\mathrm{Start}$}
 \centering{
  \includegraphics[height=0.45\textwidth]{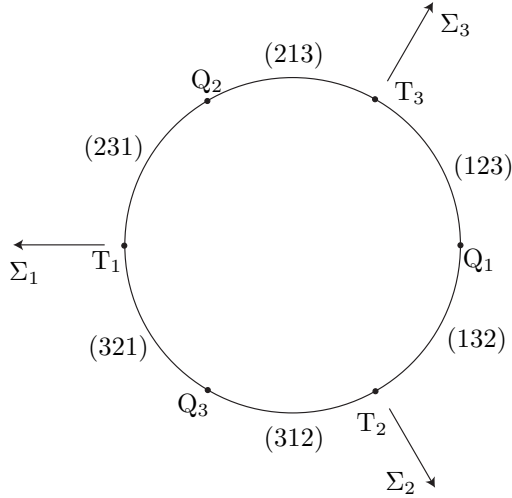}}
\caption{The Kasner circle $\mathrm{K}^{\ocircle}$ of fixed points divided
into its six equivalent sectors and the LRS fixed points
${\rm T}_\alpha$ and ${\rm Q}_\alpha$. Sector $(\alpha\beta\gamma)$ is defined by
$\Sigma_\alpha < \Sigma_\beta < \Sigma_\gamma$. The sectors are related to each other
by permutations of the spatial axes.}\label{Kasnercirc}
\end{figure}

The Bianchi type~I subset is given by $N_1 = 0$, $N_2 = 0$, $N_3 = 0$
and $\Omega = 1 -\Sigma^2 \geq 0$; since $N_1$, $N_2$, $N_3$
vanish, we denote this subset by $\mathcal{B}_\emptyset$,
cf.~Figure~\ref{contraction}. The \textit{vacuum subset} consists of
a circle of fixed points---\textit{the Kasner circle}
$\mathrm{K}^{\ocircle}$, which is characterized by $\Sigma^2=1$. It
is common to represent different points on $\mathrm{K}^{\ocircle}$
in terms of the Kasner exponents $p_\alpha$,
\begin{equation} \label{Kcircle}
(\Sigma_1,\Sigma_2,\Sigma_3) =
(3p_1-1,3p_2-1,3p_3-1)\:;\quad\qquad p_1+p_2+p_3=1\:, \quad p_1^2 +
p_2^2 + p_3^2 = 1\:;
\end{equation}
each fixed point on $\mathrm{K}^{\ocircle}$ represents a Kasner
solution (Kasner metric) with the corresponding exponents. The
Kasner circle is divided into six equivalent sectors, denoted
by permutations of the triple $(123)$, where sector
$(\alpha\beta\gamma)$ is characterized by $p_\alpha < p_\beta <
p_\gamma$, see Figure~\ref{Kasnercirc}. The boundaries of the
sectors are six special points that are associated with
solutions that are locally rotationally symmetric (LRS):
$\mathrm{Q}_\alpha$ are given by
$(\Sigma_\alpha, \Sigma_\beta, \Sigma_\gamma) = ({-2},1,1)$ or
$(p_\alpha,p_\beta,p_\gamma) =
(-\frac{1}{3},\frac{2}{3},\frac{2}{3})$ and yield the three
equivalent LRS solutions whose intrinsic geometry is non-flat;
the \textit{Taub points} $\mathrm{T}_\alpha$ are given by
$(\Sigma_\alpha, \Sigma_\beta, \Sigma_\gamma) = (2,{-}1,{-}1)$
or $(p_\alpha,p_\beta,p_\gamma) = (1,0,0)$ and correspond to
the flat LRS solutions---the Taub representation of Minkowski
spacetime.

The quantity $p_1 p_2 p_3$ (or, equivalently, $\Sigma_1 \Sigma_2
\Sigma_3 = 2 + 27 p_1 p_2 p_3$) is invariant under changes of
the axes and thus naturally captures the `physical essence' of a
solution independent of the chosen frame; however, it is typically
replaced by the \textit{Kasner parameter} $u$ through
\begin{equation}
p_1 p_2 p_3 =  \frac{-u^2(1+u)^2}{(1+u+u^2)^3}\, ,\qquad
\text{where}\quad u\in [1,\infty]\:.
\end{equation}
The Kasner parameter $u$ parameterizes the Kasner exponents uniquely
(up to the permutation symmetry); we have
\begin{equation}\label{ueq}
p_\alpha=\frac{-u}{1+u+u^2}\, ,\qquad p_\beta=\frac{1+u}{1+u+u^2}\,
,\qquad p_\gamma = \frac{u(1+u)}{1+u+u^2}\:,
\end{equation}
for sector $(\alpha\beta\gamma)$ of $\mathrm{K}^{\ocircle}$, where
$u\in(1,\infty)$.
Therefore, each point on sector $(\alpha\beta\gamma)$ is represented
by a unique value of $u \in (1,\infty)$.
At the boundary points of sector $(\alpha\beta\gamma)$,
which are $\mathrm{Q}_\alpha$ and $\mathrm{T}_\gamma$, the Kasner
parameter is $u=1$ and $u=\infty$, respectively.
Permuting $(\alpha\beta\gamma)$ yields a
physically equivalent state on a different sector; accordingly, each
$u\in(1,\infty)$ represents an equivalence class of six points on
$\mathrm{K}^\ocircle$. In contrast, $u=1$ describes the three points
$\{\mathrm{Q}_1, \mathrm{Q}_2,\mathrm{Q}_3\}$;
$u=\infty$ yields $\{\mathrm{T}_1, \mathrm{T}_2,\mathrm{T}_3\}$.

While the Bianchi type~I vacuum subset coincides with the
Kasner circle $\mathrm{K}^\ocircle$, which is given by
$\Sigma^2 = 1$, the Bianchi type~I perfect fluid subset is the
set $1- \Omega = \Sigma^2 < 1$. From~\eqref{Omegaeq} it is
straightforward to deduce that there exists a central fixed
point, the \textit{Friedmann fixed point} $\mathrm{F}$, given
by $\Sigma_\alpha=0$ $\forall \alpha$, which corresponds to the
isotropic Friedmann-Robertson-Walker (FRW) solution. Solutions
with $0<\Sigma^2 <1$ are given by radial straight lines
originating from $\mathrm{K}^\ocircle$ and ending at
$\mathrm{F}$. These results rely on the assumption $w < 1$,
see~\eqref{wassum}.

\subsection*{The Bianchi type II subset}

Let us consider the Bianchi type~II subset $\mathcal{B}_{N_\gamma}$
given by $N_\alpha = N_\beta = 0$, $N_\gamma > 0$. In this case, the
$\gamma$-direction is singled out, while there exists a discrete
symmetry associated with the interchange of the $\alpha$- and
$\beta$-direction. The LRS subset $\Sigma_\alpha = \Sigma_\beta$ is
a subset of codimension one which divides the state space into two
equivalent parts, the subsets $\{\Sigma_\alpha >\Sigma_\beta\}$ and
$\{\Sigma_\alpha < \Sigma_\beta\}$ (related by permuting the
$\alpha$- and $\beta$-axes).

Since the past singularity is of particular interest in our
considerations, it is convenient to choose the time
direction towards the past, i.e., to use a reversed time variable
$\tau_-$ according to
\begin{equation}
\tau_-=-\tau
\end{equation}
which we do in the remainder of this section; accordingly,
approach to the past singularity means $\tau_- \rightarrow
\infty$. (In Section~\ref{maps} we will see that vacuum type~II
orbits are the building blocks for the Mixmaster map and the
closely related Kasner map (BKL map). It is a well established
convention that forward iterations of these maps are directed
towards the singularity. To agree with this convention the use
of a past-directed time variable in a discussion of Bianchi
type~II models thus suggests itself.)

On $\mathcal{B}_{N_\gamma}$, the Gauss constraint $\Sigma^2 +
\sfrac{1}{12} N_\gamma^2 + \Omega =1$ can be used to replace
$N_\gamma$ by $\Omega$ as a dependent variable. The
system~\eqref{IXeq} thus becomes
\begin{equation}\label{IIeq}
\frac{d\Sigma_{\alpha/\beta}}{d\tau_-} =
(2-q) \Sigma_{\alpha/\beta} +{}^3\!S_{\alpha/\beta} \,,
\quad
\frac{d\Sigma_\gamma}{d\tau_-} = (2-q) \Sigma_{\gamma} + {}^3\!S_{\gamma}\,,
\quad
\frac{d\Omega}{d\tau_-} = -\Omega \left[ 2 q -(1+3 w) \right]\,,
\end{equation}
where $q = 2 \Sigma^2 + \sfrac{1}{2} (1+3 w) \Omega$ and
${}^3\!S_{\alpha/\beta} = -4 (1-\Sigma^2-\Omega)$,
${}^3\!S_{\gamma} = 8 (1-\Sigma^2-\Omega)$; we have $\Sigma^2 + \Omega < 1$.

\begin{figure}[ht]
\psfrag{a}[cc][cc]{$\Sigma_1$} \psfrag{b}[cc][cc]{$\Sigma_2$}
\psfrag{c}[cc][cc]{$\Sigma_3$} \psfrag{k}[cc][cc]{$\mathrm{T}_1$}
\psfrag{l}[cc][cc]{$\mathrm{T}_2$}
\psfrag{m}[cc][cc]{$\mathrm{T}_3$}
\psfrag{n}[cc][cc]{$\mathrm{Q}_1$}
\psfrag{o}[cc][cc]{$\mathrm{Q}_2$}
\psfrag{p}[cc][cc]{$\mathrm{Q}_3$}
\psfrag{s}[cc][cc]{$\mathrm{M}_1$}
 \centering{
  \includegraphics[height=0.45\textwidth]{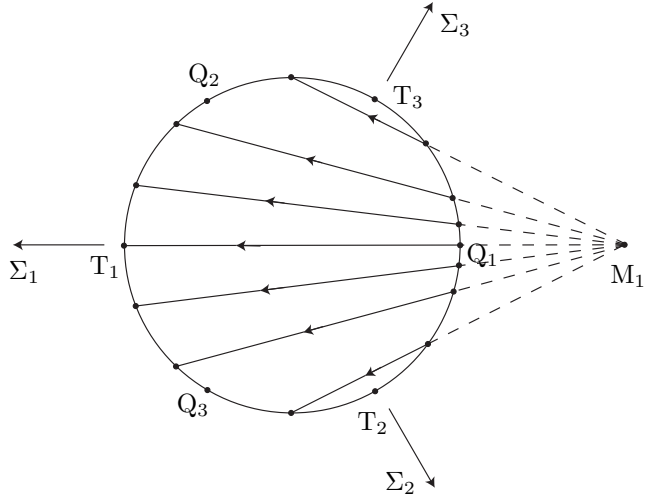}}
\caption{The type II transitions $\mathcal{T}_{N_1}$ on the
$\mathcal{B}_{N_1}$ subset; by definition, $N_1 \neq 0$ along $\mathcal{T}_{N_1}$,
while $N_2 = N_3 = 0$. The projections of these transition
onto $(\Sigma_1, \Sigma_2,\Sigma_3)$-space are straight lines, which
possess a common focal point $\mathrm{M}_1$ characterized by
$(\Sigma_1,\Sigma_2,\Sigma_3)= (-4,2,2)$.
The transitions $\mathcal{T}_{N_2}$, $\mathcal{T}_{N_3}$ on the
subsets $\mathcal{B}_{N_2}$, $\mathcal{B}_{N_3}$ are obtained by
permutations of the axes, see Figure~\ref{alltypeII}.
The arrows indicate the direction of time towards the past.}\label{typeII}
\end{figure}

Let us first consider the vacuum subset
$\mathcal{B}_{N_\gamma}^{\mathrm{vac.}}$, i.e., $\Omega = 0$.
There do not exist any fixed points in the type~II vacuum
subset $\mathcal{B}_{N_\gamma}^{\mathrm{vac.}}$, but the
boundary of the vacuum subset coincides with the Kasner circle
$\mathrm{K}^\ocircle$. The orbits of~\eqref{IIeq} form a family
of straight lines in $\mathcal{B}_{N_\gamma}^{\mathrm{vac.}}$,
where each orbit connects one fixed point on
$\mathrm{K}^\ocircle$ with another fixed point on
$\mathrm{K}^\ocircle$; hence each orbit is heteroclinic~\cite{waiell97}.
Following the nomenclature of~\cite{heietal07} we call these
heteroclinic orbits Bianchi type~II \textit{transitions}, because
each orbit can be viewed as representing a transition from one
Kasner state to another. We denote these transitions by
$\mathcal{T}_{N_\gamma}$, where each $\mathcal{T}_{N_\gamma}$
emanates from $(\gamma\alpha\beta) \cup \mathrm{Q}_\gamma \cup
(\gamma\beta\alpha)$. If the initial point is a point of sector
$(\gamma \alpha \beta)$, then the final point is a point of
$(\alpha \gamma \beta)\cup \{\mathrm{Q}_\alpha\} \cup (\alpha
\beta \gamma)$; interchanging $\alpha$ and $\beta$ yields the
transitions emanating from $(\gamma\beta\alpha)$; if the
initial point is $\mathrm{Q}_\gamma$, the final point is
$\mathrm{T}_\gamma$; the points $\mathrm{T}_\alpha$ and
$\mathrm{T}_\beta$ are not connected with any other fixed point
(they are `fixed points' under the present `type II map'), see
Figure~\ref{typeII}.

Let $(\Sigma_\alpha^{\mathrm{i}},\Sigma_\beta^{\mathrm{i}},
\Sigma_\gamma^{\mathrm{i}}) = (3 p_\alpha^{\mathrm{i}} -1,3
p_\beta^{\mathrm{i}}-1, 3p_\gamma^{\mathrm{i}}-1)$ denote the
initial fixed point on $\mathrm{K}^\ocircle$; this point can be
represented in terms of the Kasner parameter $u =
u^{\mathrm{i}}$ by using~\eqref{ueq}. The orbit (transition)
emanating from this fixed point is given in terms of an
auxiliary function $\eta = \eta(\tau_-)$ by
\begin{subequations}\label{Zeqs}
\begin{equation}
\Sigma_{\alpha/\beta} = 2 [ 1 - \eta] + \eta\, \Sigma^{\mathrm{i}}_{\alpha/\beta} \:,
\qquad
\Sigma_{\gamma} = -4 [ 1 - \eta] + \eta \,\Sigma^{\mathrm{i}}_{\gamma}\:,
\end{equation}
where $\eta$ is determined by the equation
\begin{equation}
\frac{d\eta}{d\tau_-} = 2 (1 -\Sigma^2) \eta
\qquad\text{with}\quad (1-\Sigma^{2}) = \frac{3}{g} (g - \eta)(\eta -1)
\quad\text{and}\quad
g = \frac{1+u+u^2}{1-u+u^2}
\end{equation}
\end{subequations}
and the conditions that $\lim_{\tau_-\rightarrow -\infty} \eta
=1$ and $\lim_{\tau_-\rightarrow +\infty} \eta = g$. The
quantity $g$ is in the interval $(1,3)$ for $u\in (1,\infty)$;
$u =1$ corresponds to $g = 3$ and describes the orbit
$\mathrm{Q}_\gamma \rightarrow \mathrm{T}_\gamma$; $u=\infty$
corresponds to $g=1$ and describes the `isolated' points
$\mathrm{T}_{\alpha}$, $\mathrm{T}_{\beta}$. Since $\eta$
increases from $1$ to $g$ as $\tau_-$ goes from ${-\infty}$ to
${+\infty}$, $g$ is called the growth factor~\cite{heietal07}.
In Section~\ref{maps}, the transitions
$\mathcal{T}_{N_\gamma}$, as represented by~\eqref{Zeqs}, will
appear as the building blocks for the Mixmaster/Kasner map.

While there do not exist any fixed points in the vacuum subset
of $\mathcal{B}_{N_\gamma}$, there exists one fixed point in
$\mathcal{B}_{N_\gamma}$ with $\Omega >0$, the
\textit{Collins-Stewart fixed point} $\mathrm{CS}_\gamma$,
which corresponds to one representation of the LRS solutions
found by Collins and Stewart~\cite{colste71}.
$\mathrm{CS}_\gamma$ is given by
$(\Sigma_\alpha,\Sigma_\beta,\Sigma_\gamma) = \sfrac{1}{8}(1+3
w) (1,1,-2)$ and $\Omega = 1 - \sfrac{1}{16} (1+3 w)$ (which
yields $N_\gamma = \sfrac{3}{4} \sqrt{1-w}\sqrt{1+3 w}\,$).
The fixed point $\mathrm{CS}_\gamma$ is the source (w.r.t.
$\tau_-$) for all orbits in $\mathcal{B}_{N_\gamma}$ with
$\Omega > 0$. In the limit $\tau_- \rightarrow \infty$ all
solutions in $\mathcal{B}_{N_\gamma} \backslash \{\mathrm{CS}_\gamma\}$
converge to fixed points on the Bianchi type~I
boundary of $\mathcal{B}_{N_\gamma}$: There exists one orbit
(which corresponds to an LRS solution) that converges to
$\mathrm{F}$ as $\tau_-\rightarrow \infty$; every other orbit
converges to a fixed point on $(\alpha \gamma \beta)\cup
\{\mathrm{Q}_\alpha\} \cup (\alpha \beta \gamma)$ or $(\beta
\gamma \alpha) \cup \{\mathrm{Q}_\beta\} \cup (\beta \alpha
\gamma)$ on $\mathrm{K}^\ocircle$, or to $\mathrm{T}_\gamma$
(in the LRS case). For a detailed discussion of these results
see~\cite{waiell97}.

\subsection*{The Bianchi type VII${}_{\bm{0}}$ subset}

In anticipation of Theorem~\ref{rinthm} which implies that
generic orbits of Bianchi type~IX do not have
$\alpha$-limit%
\footnote{For a dynamical system on a state space $X$,
  the $\alpha$-limit
  set $\alpha(x)$ of a point $x\in X$ is defined as the set of
  all accumulation points towards the past (i.e., as $\tau\rightarrow -\infty$)
  of the orbit $\gamma(\tau)$ through $x$. The simplest
  examples of $\alpha$-limit sets are fixed points and periodic orbits.}
points (w.r.t.\
the standard future directed time variable $\tau$) on any of
the Bianchi type~$\mathrm{VII}_0$ subsets
$\mathcal{B}_{N_1N_2}$, $\mathcal{B}_{N_2N_3}$,
$\mathcal{B}_{N_3N_1}$, we refrain from giving a detailed
discussion of these subspaces here. (However, note that in
order to \textit{prove} Theorem~\ref{rinthm}, a detailed
understanding of solutions of Bianchi type~$\mathrm{VII}_0$ is
essential; in fact, in the proof of Theorem~\ref{rinthm}
Bianchi type~$\mathrm{VII}_0$ is ubiquitous; we refer
to~\cite{rin01} and~\cite{heiuggproof}.) In the present context
it suffices to note that on each subset $\mathcal{B}_{N_\alpha
N_\beta}$ there exists a line of fixed points
$\mathrm{TL}_\gamma$ given by $(\Sigma_\alpha,\Sigma_\beta,
\Sigma_\gamma) = (-1,-1,2)$ and $N_\alpha = N_\beta$ (so that
$\Omega = 0$). Since $\mathrm{TL}_\gamma$ emanates from the
point $\mathrm{T}_\gamma \in \mathrm{K}^\ocircle$ we call it
the `Taub line.' Like $\mathrm{T}_\gamma$ itself, each of fixed
points on $\mathrm{TL}_\gamma$ is associated with a
representation of Minkowski spacetime (in an LRS type
$\mathrm{VII}_0$ symmetry foliation).

\section{Asymptotic self-similarity}
\label{nongeneric}

In the previous section we have given the fixed points associated
with the system~\eqref{IXeq} on $\overlineBIX$.
A local dynamical systems analysis of the fixed points shows
whether or not these points attract type~IX orbits
in the limit $\tau\rightarrow -\infty$.%
\footnote{In this section we adapt to~\cite{waiell97,rin01} and
  use the normal future-directed time variable $\tau$.}
We merely state the results here and refer to~\cite{heiuggproof}
for details.

\begin{figure}[ht]
\psfrag{a}[cc][cc]{$\Sigma_1$} \psfrag{b}[cc][cc]{$\Sigma_2$}
\psfrag{c}[cc][cc]{$\Sigma_3$} \psfrag{d}[cc][cc]{$N_1$}
\psfrag{e}[cc][cc]{$N_2$} \psfrag{f}[cc][cc]{$N_2$}
\psfrag{g}[cc][cc]{$N_3$} \psfrag{h}[cc][cc]{$N_3$}
\psfrag{i}[cc][cc]{$N_1$} \psfrag{k}[cc][cc]{$\mathrm{T}_1$}
\psfrag{l}[cc][cc]{$\mathrm{T}_2$}
\psfrag{m}[cc][cc]{$\mathrm{T}_3$}
\psfrag{n}[cc][cc]{$\mathrm{Q}_1$}
\psfrag{o}[cc][cc]{$\mathrm{Q}_2$}
\psfrag{p}[cc][cc]{$\mathrm{Q}_3$}
\psfrag{s}[cc][cc]{$\mathrm{M}_1$}
\psfrag{t}[cc][cc]{$\mathrm{M}_2$}
\psfrag{u}[cc][cc]{$\mathrm{M}_3$}
\psfrag{x}[cc][cc]{$\mathrm{Start}$}
 \centering{
  \includegraphics[height=0.45\textwidth]{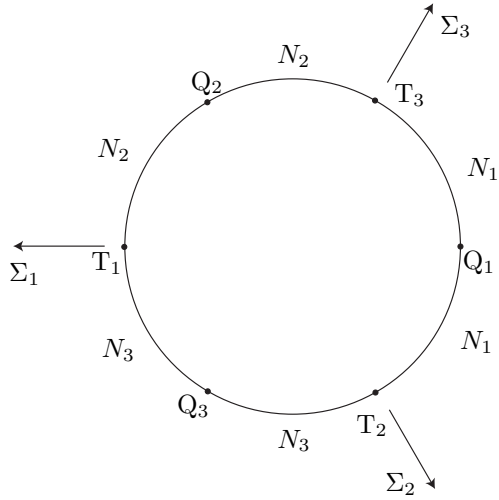}}
\caption{This figure depicts the Kasner circle $\mathrm{K}^{\ocircle}$
and the stable variables for each sector; $N_\alpha$ is the stable variable in
sectors $(\alpha\beta\gamma)$, $(\alpha\gamma\beta)$, and at the
point $\mathrm{Q}_\alpha$. Expressed in the time variable $\tau_-$,
which is directed towards the past, these variables are the unstable
modes. At a given fixed point, the associated unstable manifold
orbit is a Bianchi type~II transition,
see~Figure~\ref{typeII}.} \label{unstable}
\end{figure}

\begin{itemize}
\item[$\mathrm{K}^{\ocircle}$] Each fixed point
    $\mathrm{K}$ on $\mathrm{K}^{\ocircle}
    \backslash\{\mathrm{T}_1,\mathrm{T}_2, \mathrm{T}_3\}$
    is a
    transversally hyperbolic saddle (one stable%
\footnote{When we reverse the direction of time,
  i.e., when we use $\tau_-$ instead of $\tau$,
  we must replace `stable' by `unstable'.}
mode, see Figure~\ref{unstable}, and three unstable modes; the
stable manifold of $\mathrm{K}$ coincides with a vacuum type~II transition orbit,
see Figure~\ref{typeII}).
The Taub points $\{\mathrm{T}_1,\mathrm{T}_2,
\mathrm{T}_3\}$ are center saddles with a two-dimensional
unstable manifold and a three-dimensional center manifold,
where $\mathrm{T}_\alpha$ is excluded as an $\alpha$-limit
set on the center manifold,
see~\cite{heiuggproof}; consequently there do not exist any
type~IX solutions that converge to any of the points on
$\mathrm{K}^\ocircle$ as $\tau\rightarrow -\infty$.
\item[F\,\,] The fixed point $\mathrm{F}$ on
    $\mathcal{B}_{\emptyset}^{\mathrm{fl.}}$ is a
    hyperbolic saddle (with
    $\mathcal{B}_{\emptyset}^{\mathrm{fl.}}$ as a
    two-di\-men\-sional stable manifold and an additional
    three-di\-men\-sional unstable manifold). Accordingly,
    $\mathrm{F}$ attracts a two-parametric family of
    type~IX orbits as $\tau\rightarrow -\infty$. These
    solutions have a so-called isotropic singularity.
\item[$\mathrm{CS}_\alpha$] The fixed points
    $\mathrm{CS}_\alpha$ ($\alpha = 1,2,3$) on
    $\mathcal{B}_{N_\alpha}^{\mathrm{fl.}}$ are hyperbolic
    saddles (with $\mathcal{B}_{N_\alpha}^{\mathrm{fl.}}$
    as a three-di\-men\-sional stable manifold and an additional
    two-dimensional unstable manifold). The unstable modes
    are associated with the equations
    $N_\beta^{-1}N_\beta'\,|_{\mathrm{CS}_\alpha} =
    \sfrac{3}{4} (1+3 w)$ (for $\beta \neq \alpha$).
    Therefore, each of the fixed points
    $\mathrm{CS}_\alpha$ attracts an (equivalent)
    one-parameter set of type~IX orbits in the limit
    $\tau\rightarrow -\infty$.
\item[$\mathrm{TL}_\alpha$] Each fixed point on
    $\mathrm{TL}_\alpha$ on $\mathcal{B}_{N_\beta
    N_\gamma}^{\mathrm{vac.}}$ is a center saddle. On the
    three-dimensional center manifold (which coincides with
    $\mathcal{B}_{N_\beta
    N_\gamma}^{\,\mathrm{vac.}}$, $\alpha\neq \beta \neq
    \gamma \neq \alpha$) the point acts as a
    (non-hyperbolic) sink, see~\cite{heiuggproof}. Since
    there is a two-dimensional unstable manifold, there
    exists, for each fixed point on $\mathrm{TL}_\alpha$, a
    one-parameter family of type~IX orbits that converges
    to it as $\tau\rightarrow -\infty$; these orbits
    correspond to LRS solutions. (Conversely, generic LRS
    type~IX solutions converge to $\mathrm{TL}_\alpha$,
    see, e.g.,\cite{waiell97}.)
\end{itemize}

Collecting the results we see that the solutions whose
$\alpha$-limit is one of the fixed points form a subfamily of
measure zero of the (four-parameter) family of Bianchi type~IX
solutions. Following the nomenclature of~\cite{rin01} we thus
refer to these solutions as \textit{non-generic} solutions of
Bianchi type~IX. Alternatively, to capture the asymptotic
behavior of these solution, we use the term \textit{past
asymptotically self-similar} solutions. (Since a fixed point in
the Hubble-normalized dynamical systems formulation corresponds
to a self-similar solution, see e.g.~\cite{waiell97}, solutions
that converge to a fixed point are asymptotically
self-similar.)

The past asymptotically self-similar solutions comprise the LRS
Bianchi type~IX solutions. As seen above, generic LRS solutions
converge to $\mathrm{TL}_\alpha$ towards the past (and each solution
that converges to $\mathrm{TL}_\alpha$ is LRS), but there exist
exceptional LRS solutions that converge to $\mathrm{F}$ or
$\mathrm{CS}_\alpha$. The remaining orbits whose limit point is
either $\mathrm{F}$ or $\mathrm{CS}_\alpha$ correspond to past
asymptotically self-similar solutions that are non-LRS.
Clearly, every solution that converges to $\mathrm{F}$ or
$\mathrm{CS}_\alpha$ is a non-vacuum solution, since $\Omega \neq 0$
at $\mathrm{F}$ and $\mathrm{CS}_\alpha$.

It is natural to ask how the non-generic orbits are embedded in the
state space $\overlineBIX$. The LRS orbits form the three LRS
subsets $\mathcal{L\!R\!S}_\alpha$, which are the hyperplanes given
by the conditions $\Sigma_\beta = \Sigma_\gamma$, $N_\beta =
N_\gamma$, where $(\alpha\beta\gamma) \in \{(123),(231),(312)\}$.
The orbits whose $\alpha$-limit set is the fixed point
$\mathrm{CS}_\alpha$ (for some $\alpha$) form the set
$\mathcal{C\!S}_\alpha$ in $\BIX$; we call $\mathcal{C\!S}_\alpha$ the
Collins-Stewart manifold. The local analysis of the fixed point
$\mathrm{CS}_\alpha$ and the regularity of the dynamical
system~\eqref{IXeq} imply that the Collins-Stewart manifold
$\mathcal{C\!S}_\alpha$ is a two-dimensional surface; it can be viewed
as a two-dimensional manifold with boundary embedded in
$\overlineBIX$ (where this boundary corresponds to an orbit in $\overlineBVII$).
Analogously, the orbits whose $\alpha$-limit set is the fixed point
$\mathrm{F}$ form the set $\mathscr{F}$ in $\BIX$, which we call the
isotropic singularity manifold, since solutions converging to $\mathrm{F}$
are those with an isotropic singularity. The isotropic singularity
manifold
$\mathscr{F}$ is a three-dimensional hypersurface; it can be viewed
as a three-dimensional manifold with boundary.

Generic Bianchi type~IX models are those that are not
asymptotic self-similar and thus constitute examples for
asymptotic self-similarity breaking; for other such examples,
see~\cite{limetal06,waietal99}. The central theme in this paper
is the past asymptotic behavior of the generic models.

\section{Facts about the Mixmaster and Kasner map}
\label{maps}

The \emph{Mixmaster attractor\/} $\mathcal{A}_{\mathrm{IX}}$
(alternatively referred to as the Bianchi type~IX attractor) is
defined to be the subset of $\overlineBIX$ given by the union
of the Bianchi type~I and~II vacuum subsets, i.e.,
$\mathcal{A}_{\mathrm{IX}} = \BI^{\mathrm{vac.}} \cup
\BII^{\mathrm{vac.}}$. Since the type~II vacuum subset consists
of three equivalent representations we obtain
\begin{equation}\label{AIXdef}
\mathcal{A}_{\mathrm{IX}} = \mathrm{K}^\ocircle \cup
\mathcal{B}_{N_1}^{\mathrm{vac.}} \cup
\mathcal{B}_{N_2}^{\mathrm{vac.}} \cup
\mathcal{B}_{N_3}^{\mathrm{vac.}} \:.
\end{equation}
In this section we investigate the structures that the
flow of dynamical system~\eqref{IXeq} induces on $\mathcal{A}_{\mathrm{IX}}$.
In particular we discuss the Mixmaster map, the Kasner
map, and the era map. To agree with the well-established convention
for these maps, the direction of time will be taken towards the past.

\subsection*{The Mixmaster, Kasner, and era maps}

\begin{figure}[th]
\psfrag{t1}[rc][cc][0.9][0]{$\mathrm{T}_1$}
\psfrag{t2}[tl][tl][0.9][0]{$\mathrm{T}_2$}
\psfrag{t3}[bl][bl][0.9][0]{$\mathrm{T}_3$}
\psfrag{q1}[lc][cc][0.9][0]{$\mathrm{Q}_1$}
\psfrag{q2}[br][br][0.9][0]{$\mathrm{Q}_2$}
\psfrag{q3}[tr][tr][0.9][0]{$\mathrm{Q}_3$}
\centering
\includegraphics[width=0.25\textwidth]{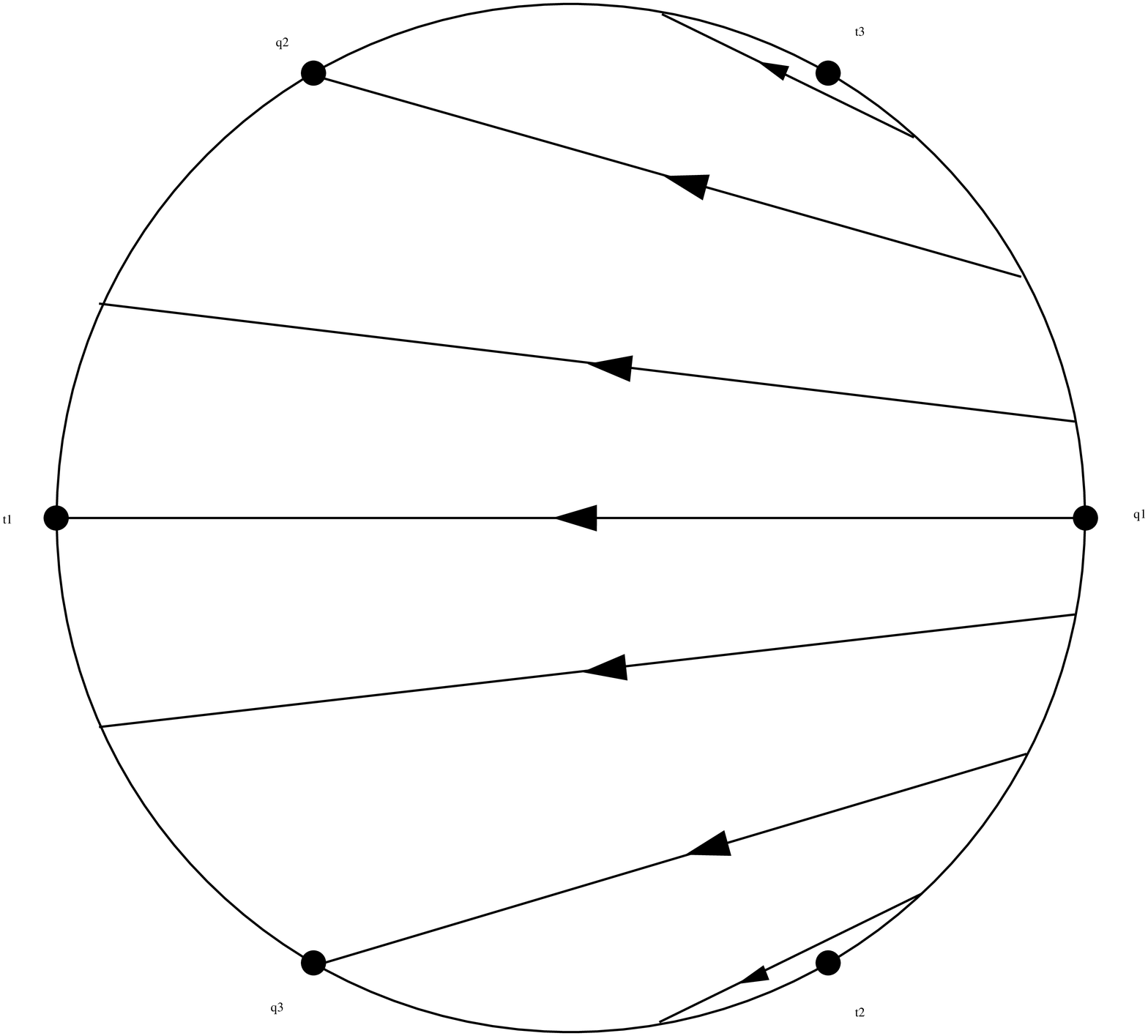}\qquad\qquad
\includegraphics[width=0.25\textwidth]{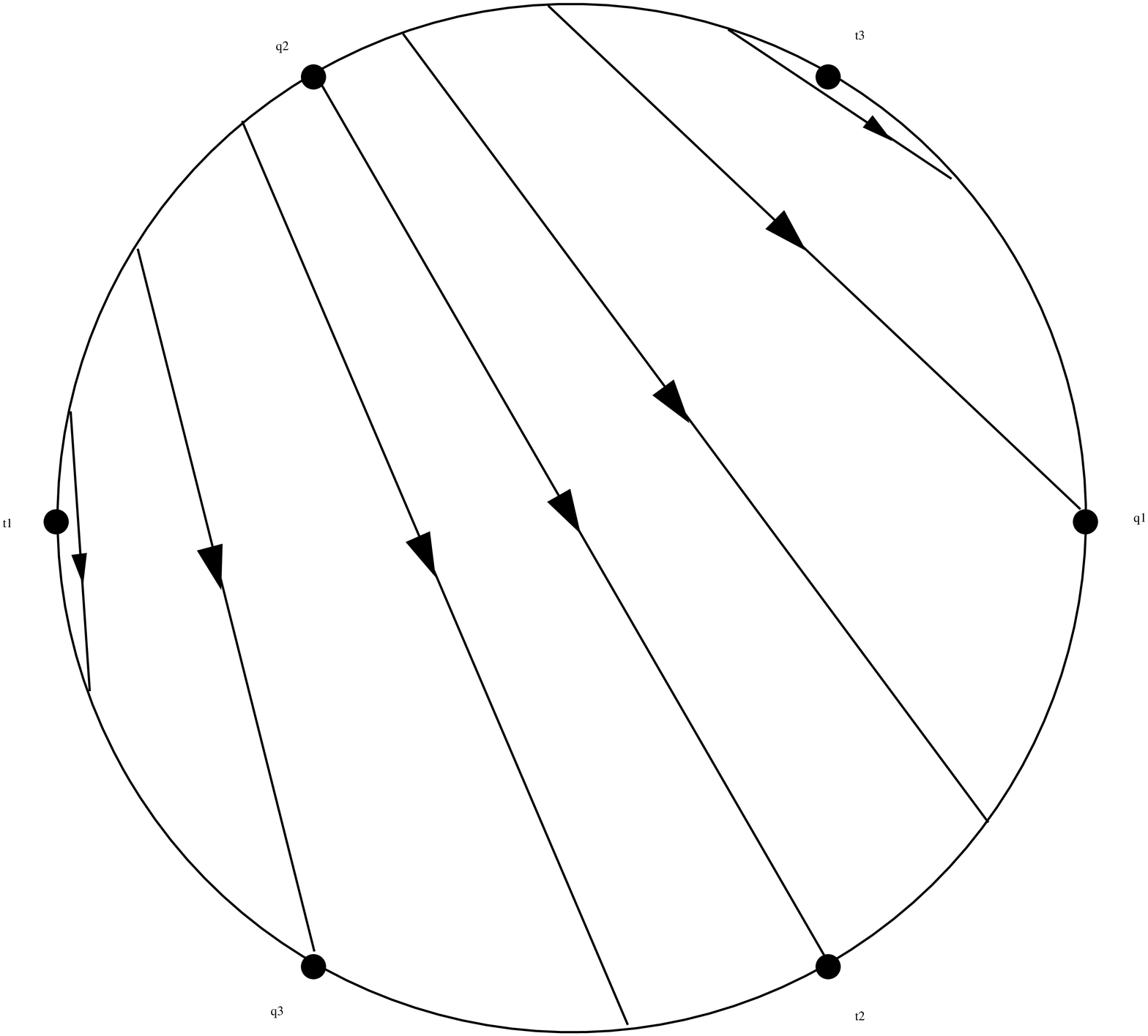}\qquad\qquad
\includegraphics[width=0.25\textwidth]{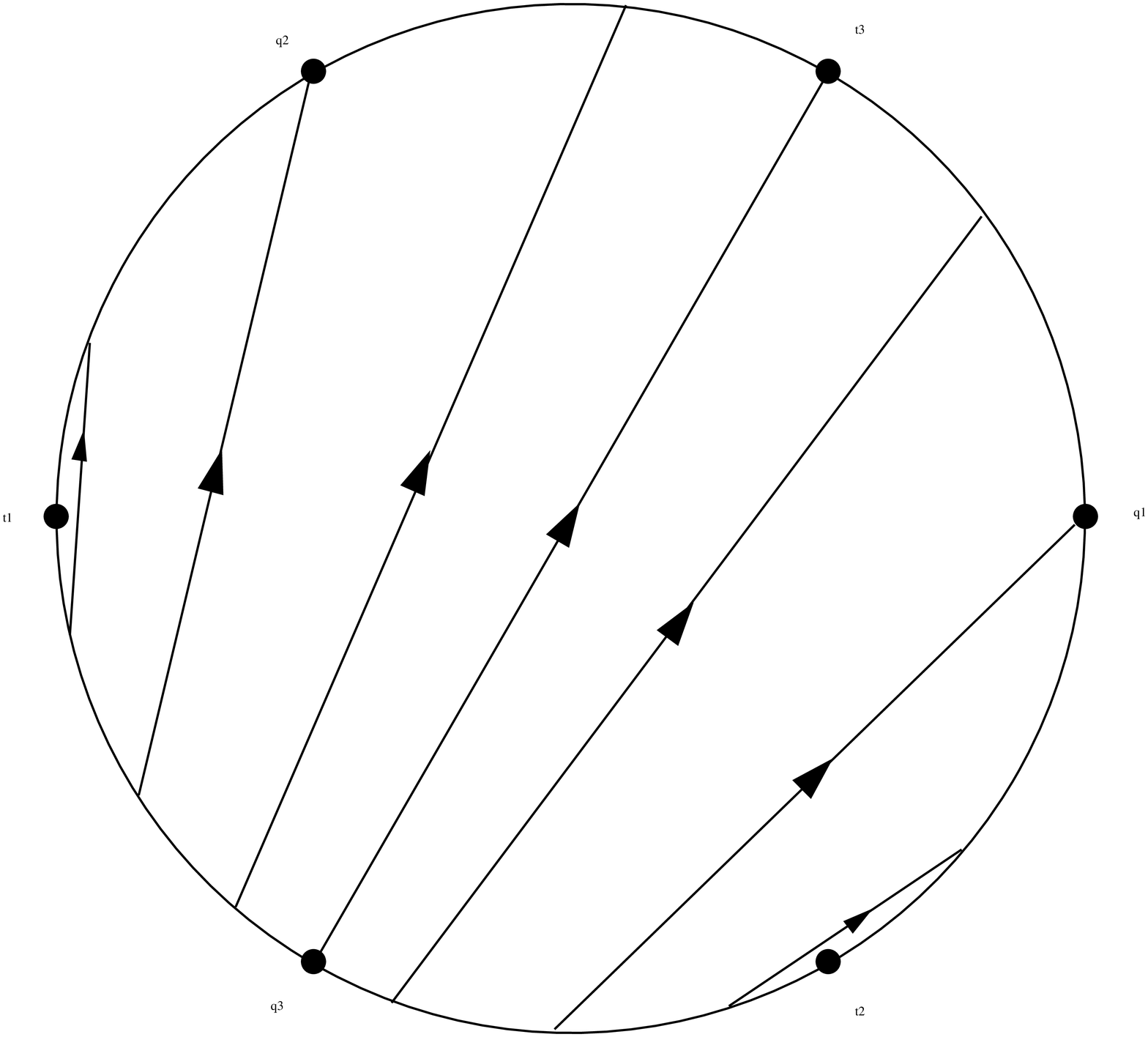}
\caption{Projection of the type~II transitions $\mathcal{T}_{N_1}$ $\mathcal{T}_{N_2}$, $\mathcal{T}_{N_3}$
on the type~II subsets
$\mathcal{B}_{N_1}^{\mathrm{vac.}}$, $\mathcal{B}_{N_2}^{\mathrm{vac.}}$, $\mathcal{B}_{N_3}^{\mathrm{vac.}}$
onto $(\Sigma_1, \Sigma_2,\Sigma_3)$-space.
Concatenation of these transition orbits yields sequences of
transitions. The arrows indicate the direction of time towards the past.}
\label{alltypeII}
\end{figure}

In Section~\ref{subsets} we have seen that
the vacuum type~II orbits, i.e., the orbits
on $\mathcal{B}_{N_\alpha}^{\mathrm{vac.}}$, $\alpha = 1,2,3$,
are heteroclinic orbits that
emanate from and converge to fixed points on the Kasner circle
$\mathrm{K}^\ocircle$, see Figures~\ref{typeII} and~\ref{alltypeII}.
In accord with Section~\ref{subsets} we refer to
these orbits as \textit{transitions}
and denote them by
$\mathcal{T}_{N_\alpha}$, $\alpha=1,2,3$.

The type~II transitions are the building blocks for the
analysis of the Mixmaster attractor. By concatenating transitions we
obtain a \textit{sequence of transitions}, also known as a
\textit{heteroclinic chain}. Since each fixed point on
$\mathrm{K}^\ocircle$ (except for the Taub points) is the
initial value for one single transition, see
Figures~\ref{unstable} and~\ref{alltypeII},
the concatenation of transitions is unique:
Each fixed point (except for the Taub points)
generates a unique sequence of transitions.
Note, however, that the `direction of time' is relevant. For
each fixed point $\mathrm{P}$ on $\mathrm{K}^\ocircle$, which
is not one of the Taub points, there exists one single
transition emanating from $\mathrm{P}$ at $\tau_- = -\infty$,
but there are two transitions converging to $\mathrm{P}$ as
$\tau_- \rightarrow \infty$. Therefore, concatenating
transitions in the reversed direction of time leads to
ambiguities. (In terms of the standard future-directed time
variable $\tau$ we have the converse statement: It is possible
to make unambiguous retrodictions, but not predictions.)

\begin{figure}[ht]
\psfrag{a}[cc][cc]{$\Sigma_1$} \psfrag{b}[cc][cc]{$\Sigma_2$}
\psfrag{c}[cc][cc]{$\Sigma_3$} \psfrag{k}[cc][cc]{$\mathrm{T}_1$}
\psfrag{l}[cc][cc]{$\mathrm{T}_2$}
\psfrag{m}[cc][cc]{$\mathrm{T}_3$}
\psfrag{n}[cc][cc]{$\mathrm{Q}_1$}
\psfrag{o}[cc][cc]{$\mathrm{Q}_2$}
\psfrag{p}[cc][cc]{$\mathrm{Q}_3$}
\psfrag{s}[cc][cc]{$\mathrm{M}_1$}
\psfrag{t}[cc][cc]{$\mathrm{M}_2$}
\psfrag{u}[cc][cc]{$\mathrm{M}_3$}
\psfrag{x}[cc][cc]{$\mathrm{Start}$}
 \centering{
  \includegraphics[height=0.45\textwidth]{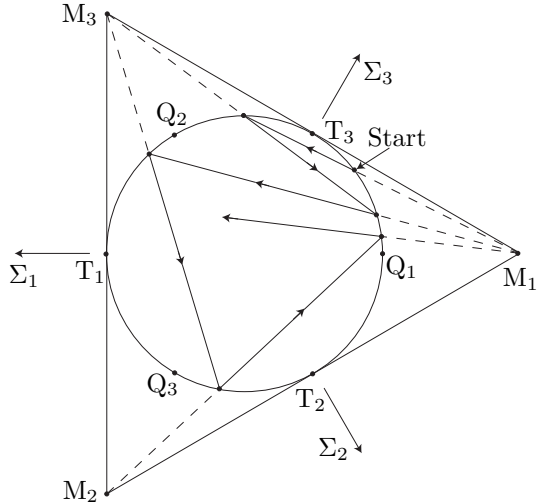}}
\caption{Concatenating type~II transition orbits
we obtain sequences of transitions---heteroclinic chains. The discrete map governing
the associated sequence of fixed points on $\mathrm{K}^\ocircle$
is the Mixmaster map.
The arrows indicate the direction of time towards the past.}
\label{Mixmaster}
\end{figure}

Let $l=0,1,2,\ldots$ and let $\mathrm{P}_l \in
\mathrm{K}^\ocircle$ denote the initial point of the
$l$\raisebox{0.7ex}{\small th} transition ($\mathrm{P}_l$ is
also the end point of the $(l-1)$\raisebox{0.7ex}{\small th}
transition). We refer to the sequence
$(\mathrm{P}_l)_{l\in\mathbb{N}}$ of Kasner fixed points, which
is induced by the sequence of transitions, as being generated
by the~\textit{Mixmaster map}. The Mixmaster map can be
visualized by a map in $(\Sigma_1, \Sigma_2,\Sigma_3)$-space,
obtained by inscribing $\mathrm{K}^{\ocircle}$ in a triangle
with corners at $(\Sigma_1,\Sigma_2,\Sigma_3)= (-4,2,2)$ and
cyclic permutations, from which the (projections of the)
transition orbits `originate' as straight lines; see
Figure~\ref{Mixmaster}.

Let the initial Kasner state of a transition be represented by
the Kasner parameter $u = u^{\mathrm{i}}$, where we assume
$u^{\mathrm{i}} < \infty$, since neither of the Taub points
$\mathrm{T}_1$, $\mathrm{T}_2$, $\mathrm{T}_3$ can be the
initial value for a transition. Inserting~\eqref{Kcircle}
and~\eqref{ueq} into~\eqref{Zeqs} we find that a transition
maps the Kasner parameter $u^{\mathrm{i}}$ to the parameter
$u^{\mathrm{f}}$, where
\begin{equation}\label{BKLMap}
u^{\mathrm{f}}  \:= \: \left\{\begin{array}{ll}
u^{\mathrm{i}} - 1 & \qquad \text{if}\quad u^{\mathrm{i}} \in [2, \infty)\:, \\[1ex]
(u^{\mathrm{i}} - 1)^{-1} &\qquad  \text{if} \quad u^{\mathrm{i}} \in [1,2]\:.
\end{array}\right.
\end{equation}
The information contained in this \emph{Kasner map} suffices to
represent the collection of all transition orbits (as a whole).
However, for each particular mapping $u^{\mathrm{i}} \mapsto
u^{\mathrm{f}}$, there exist six (equivalent) associated
transitions;%
\footnote{In the exceptional case $u^{\mathrm{i}} = 1$ there exist
  only three (equivalent) associated transitions: $\mathrm{Q}_1\rightarrow \mathrm{T}_1$,
  $\mathrm{Q}_2\rightarrow \mathrm{T}_2$, $\mathrm{Q}_3\rightarrow \mathrm{T}_3$.}
this is simply because $u^{\mathrm{i}}$ characterizes the initial
Kasner fixed point on $\mathrm{K}^\ocircle$ only up to permutations
of the axes. Hence, in order to reconstruct a particular transition
from~\eqref{BKLMap}, we supplement~\eqref{BKLMap} with information
about the initial sector of the transition, which determines the
position of the axes.

In terms of the Kasner parameter, a sequence of transitions
corresponds to an iteration of~\eqref{BKLMap}. Let $l=0,1,2,\ldots$
and let $u_l$ denote the initial Kasner state of the
$l$\raisebox{0.7ex}{\small th} transition. This transition maps
$u_l$ to $u_{l+1}$, i.e.,
\begin{equation}\label{Kasnermap}
u_l \:\,\xrightarrow{\;\text{$l$\raisebox{0.5ex}{th} transition}\;}\:\, u_{l+1}\::
\qquad\quad
u_{l+1} \:= \: \left\{\begin{array}{ll}
u_l - 1 & \qquad \text{if}\quad u_l \in[2,\infty)\,, \\[1ex]
(u_l - 1)^{-1} &\qquad  \text{if} \quad u_l \in [1,2]\,.
\end{array}\right.
\end{equation}
We refer to this map as the (iterated) \textit{Kasner map} (which is
also known as the BKL map~\cite{bkl70}). Since each value of the
Kasner parameter $u \in(1,\infty)$ represents an equivalence class
of six Kasner fixed points, the Kasner map can be regarded as the
map induced by the Mixmaster map on these equivalence classes via
the equivalence relation.

In a sequence $(u_l)_{l=0,1,2,\ldots}$ that is generated by the
Kasner map~\eqref{Kasnermap}, each Kasner state $u_l$ is called
an \textit{epoch}. Every sequence $(u_l)_{l=0,1,2,\ldots}$
possess a natural partition into pieces (which contain a finite
number of epochs each) where the Kasner parameter is
monotonically decreasing according to the simple rule $u_l
\mapsto u_{l+1} = u_l - 1$; these pieces are called
\textit{eras}~\cite{bkl70}.
An era 
begins with a maximal value $u_{\lin}$ of the Kasner parameter
(where $u_{\lin}$ is generated from $u_{\lin-1}$ by $u_{\lin} =
[u_{\lin-1}-1]^{-1}$), continues with a sequence of Kasner
parameters obtained via $u_l \mapsto u_{l+1} = u_l - 1$, and
ends with a minimal value $u_{\lout}$ that satisfies $1 <
u_{\lout} < 2$, so that $u_{\lout+1} = [u_{\lout}-1]^{-1}$
begins a new era.
\begin{equation}\label{phases}
\underbrace{6.29 \rightarrow 5.29 \rightarrow 4.29 \rightarrow 3.29
\rightarrow 2.29 \rightarrow 1.29}_{\text{\scriptsize era}} \rightarrow
\underbrace{3.45 \rightarrow 2.45 \rightarrow 1.45}_{\text{\scriptsize era}}  \rightarrow
\underbrace{2.23 \rightarrow 1.23}_{\text{\scriptsize era}}  \rightarrow
\underbrace{4.33 \rightarrow \ldots}_{\text{\scriptsize era}}
\end{equation}

Let us denote the initial ($=$ maximal) value of the Kasner
parameter $u$ in era number $s$ (where $s = 0,1,2,\ldots$) by
$\u_s$. Following~\cite{bkl70,khaetal85}
we decompose $\u_s$ into its integer part $k_s = [\u_s]$ and
its fractional part $x_s = \{\u_s\}$, i.e.,
\begin{equation}\label{usdecomp}
  \u_s = k_s + x_s\:,\qquad\qquad  \text{where}\quad k_s= [\u_s]\:, \quad
  x_s= \{ \u_s\}\:.
\end{equation}
The number $k_s$ represents the (discrete) length of era $s$, which
is simply the number of Kasner epochs it contains. The final ($=$
minimal) value of the Kasner parameter in era $s$ is given by $1 +
x_s$, which implies that era number $(s+1)$ begins with
\begin{equation*}
  \u_{s+1} = \frac{1}{x_s} = \frac{1}{\{\u_s\}}\:.
\end{equation*}
The map $\u_s \mapsto \u_{s+1}$ is (a variant of) the so-called
\textit{era map}; starting from $\u_0 = u_0$ it recursively
determines $\u_s$, $s=0,1,2,\ldots$, and thereby the complete Kasner
sequence $(u_l)_{l=0,1,\ldots}$.

The era map admits a straightforward interpretation in terms of continued fractions.
Consider the continued fraction representation of the initial value, i.e.,
\begin{equation}
  \u_0 =  k_0 + \cfrac{1}{k_1 + \cfrac{1}{k_2 + \dotsb}} = [k_0; k_1,k_2,k_3,\dotsc]\:.
\end{equation}
The fractional part of $\u_0$ is $x_0 = [0;k_1,k_2,k_3,\dotsc]$;
since $\u_1$ is the reciprocal of $x_0$ we find
\begin{equation}
  \u_1 = [k_1; k_2,k_3,k_4,\dotsc]\:.
\end{equation}
Therefore, the era map is simply a shift to the left in
the continued fraction expansion,
\begin{equation}
  \u_s = [k_s; k_{s+1}, k_{s+2}, \dotsc] \:\mapsto\:
  \u_{s+1} = [k_{s+1}; k_{s+2}, k_{s+3},\dotsc]\:.
\end{equation}

The properties of the Kasner sequence depend on the initial value $\u_0 = u_0$.
\begin{itemize}
\item[(i)] If and only if the initial Kasner parameter $u_0$ is a rational number,
  i.e., if and only if $u_0 \in \mathbb{Q}$, then its continued fraction
  representation is finite, i.e.,
  \begin{subequations}\label{cfki}
  \begin{equation}\label{terminalki}
    u_0 = [ k_0; k_1, k_2, \dotsc, k_n ]\:,
    \tag{\ref{cfki}i}
  \end{equation}
  where $k_n > 1$. Therefore, there exists only a finite number of eras
  (where the last one begins with $\u_n = k_n$), and the Kasner
  sequence is finite. At the end of era number $n$, the Kasner
  parameter reaches $u=1$, which subsequently terminates the recursion~\eqref{Kasnermap}.
  Since $\mathbb{Q}$ is a set of measure
  zero in $\mathbb{R}$, this case is non-generic.
\item[(ii)] A quadratic irrational (quadratic surd) is an
  algebraic number of degree $2$, i.e., an
  irrational solution of a quadratic equation with integer
  coefficients.
  If and only if the initial Kasner parameter $u_0$ is
  a quadratic irrational,
  i.e., if and only if $u_0 = q_1 + \sqrt{q_2}$,
  where $q_1 \in \mathbb{Q}$ and $q_2 \in \mathbb{Q}$ is not a perfect square
  (i.e., $\sqrt{q_2} \not\in \mathbb{Q}$),
  then its continued fraction representation is periodic, i.e.,
  \begin{equation}\label{periodicki}
    u_0 = \big[ k_0; k_1, \dotsc,k_n, (\bar{k}_{1}, \dotsc, \bar{k}_{\bar{n}})\big]\:;
    \tag{\ref{cfki}ii}
  \end{equation}
  the notation is such that the
  part in parenthesis, i.e., $(\bar{k}_{1}, \dotsc, \bar{k}_{\bar{n}})$,
  is repeated ad infinitum.%
  \footnote{If $u_0 = q_1 + \sqrt{q_2}> 1$ is a quadratic irrational such that
    $q_1 - \sqrt{q_2} \in (-1,0)$, then $u_0 = \big[ (\bar{k}_{1}, \dotsc, \bar{k}_{\bar{n}})\big]$,
    i.e., the continued fraction is purely periodic without any preperiod.}
  Consequently, the era map becomes periodic (after the $n$\raisebox{0.7ex}{\small th} era),
  and we thus obtain
  a periodic sequence of eras and a periodic Kasner sequence $(u_l)_{l=0,1,2,\ldots}$.
  It is straightforward to see that while the period of the era sequence is $\bar{n}$,
  the period of the Kasner sequence is $(\bar{k}_1 + \cdots + \bar{k}_{\bar{n}})$;
  see the examples below.
  Since the set of algebraic numbers of degree two (or equivalently the set of
  equations with integer
  coefficients---it is a subset of $\mathbb{N}^3$) is a countable set,
  case (ii) is also non-generic.
\item[(iii)] An irrational number is called badly approximable if its Markov constant%
  \footnote{For $x\in\mathbb{R}$, let $\|x\|$ denote the distance from $x$
    to the nearest integer, i.e., $\| x\| = \min_{n\in\mathbb{Z}}|x-n|$.
    The Markov constant $M(x)$ of a number $x \in \mathbb{R}\backslash\mathbb{Q}$
    is defined as $M(x)^{-1} = \liminf_{\mathbb{N}\ni n\rightarrow \infty}  n \,\|n x \|$,
    see~\cite{tri93}.
    It is known that $M(x) \geq \sqrt{5}$ for all $x\in\mathbb{R}\backslash\mathbb{Q}$.}
  is finite. If and only if $u_0$ is badly approximable, then the coefficients (partial quotients)
  in its continued fraction representation are bounded, i.e.,
  \begin{equation}\label{boundedki}
    u_0 = \big[ k_0; k_1, k_2, k_3, \dotsc \big] \qquad \text{with}\quad k_i \leq K\:\, \forall i
   \tag{\ref{cfki}iii}
  \end{equation}
  for some positive constant $K$.
  Consequently, the sequence of eras and the Kasner sequence $(u_l)_{l=0,1,2,\ldots}$ are bounded,
  i.e., $u_l \leq K$ $\forall l$.
  Obviously, case (ii) is a subcase of case (iii). (Note, however, that there is
  probably no relationship between case (iii) and
  algebraic numbers of degree greater than~$2$, since
  it is expected that these numbers are well approximable.)
  The set of badly approximable numbers are a set of Lebesgue measure zero,
  hence this case is non-generic.
\item[(iv)] If and only if the initial Kasner parameter
    $u_0$ is a well approximable irrational number, then
    the partial quotients $k_i$ in the continued fraction
    representation
  \begin{equation}\label{wellki}
    u_0 = \big[ k_0; k_1, k_2 , k_3, \dotsc \big]
    \tag{\ref{cfki}iv}
  \end{equation}
  \end{subequations}
  are unbounded (and we can construct a diverging subsequence from the sequence
  of partial quotients $(k_i)_{i\in\mathbb{N}}$).
  This is the generic case, and hence generically the Kasner sequence
  $(u_l)_{l=0,1,2,\ldots}$ is infinite and unbounded.
\end{itemize}

In terms of continued fractions, the Kasner sequence $(u_l)_{l\in\mathbb{R}}$
generated by $u_0 =  \big[ k_0; k_1, k_2 , \dotsc \big]$ is
\begin{align*}
u_0 = \mathsf{u}_0 & =
\big[ k_0; k_1, k_2,  \dotsc \big] \rightarrow  \big[ k_0 -1 ; k_1, k_2 , \dotsc \big]
\rightarrow \big[ k_0 -2 ; k_1, k_2 ,  \dotsc \big]
\rightarrow \ldots \rightarrow \big[1 ; k_1, k_2 , \dotsc \big] \\
 \rightarrow \mathsf{u}_1 & = \big[ k_1; k_2 , k_3, \dotsc \big]
 \rightarrow  \big[  k_1-1; k_2 , k_3, \dotsc \big]
\rightarrow \big[ k_1 -2 ; k_2 , k_3, \dotsc \big]
\rightarrow \ldots \rightarrow \big[1 ; k_2, k_3 , \dotsc \big] \\
 \rightarrow \mathsf{u}_2 & = \big[ k_2; k_3 , k_4, \dotsc \big]
 \rightarrow  \big[  k_2-1; k_3 , k_4, \dotsc \big]
\rightarrow \big[ k_2 -2 ; k_3 , k_4, \dotsc \big] \rightarrow \ldots
\end{align*}

Let us give some examples for periodic {\em era sequences\/} and
{\em Kasner sequences\/}. If $u_0 = [(1)] = (1+\sqrt{5})/2$,
which is the golden ratio, then $\u_s = (1+\sqrt{5})/2$ $\forall s$.
It follows that the Kasner sequence is also a sequence with period $1$,
\begin{equation*}
(u_l)_{l\in\mathbb{N}}: \sfrac{1}{2}\big(1+\sqrt{5}\big) \rightarrow
\sfrac{1}{2}\big(1+\sqrt{5}\big) \rightarrow
\sfrac{1}{2}\big(1+\sqrt{5}\big) \rightarrow
\sfrac{1}{2}\big(1+\sqrt{5}\big) \rightarrow
\sfrac{1}{2}\big(1+\sqrt{5}\big) \rightarrow
\ldots
\end{equation*}
If $u_0 = [(2)] = 1 +\sqrt{2}$, then
$\u_s = 1+\sqrt{2}$ $\forall s$; hence the era sequence is a sequence of
period $1$. However, the associated Kasner sequence has period $2$,
\begin{equation*}
(u_l)_{l\in\mathbb{N}}:
\underbrace{(1+\sqrt{2}) \rightarrow \sqrt{2}}_{\text{\scriptsize{era}}} \rightarrow
\underbrace{(1+\sqrt{2}) \rightarrow \sqrt{2}}_{\text{\scriptsize{era}}} \rightarrow
\underbrace{(1+\sqrt{2}) \rightarrow \sqrt{2}}_{\text{\scriptsize{era}}} \rightarrow
\underbrace{(1+\sqrt{2}) \rightarrow \sqrt{2}}_{\text{\scriptsize{era}}} \rightarrow
\ldots
\end{equation*}
Analogously, the initial value $u_0 = [(3)] = (3+\sqrt{13})/2$
generates an era sequence of period $1$ and an associated Kasner sequence
of period $3$.
Finally let $u_0 = [(2,4)] = 1 + \sqrt{3/2} \simeq 2.2247$.
Then the era sequence has period $2$ with $\u_n = [(2,4)] \simeq 2.2247$ for
even $n$ and $\u_n = [(4,2)] \simeq 4.4495$ for odd $n$.
The associated Kasner sequence $(u_l)_{l\in\mathbb{N}}$ has period $6$,
\begin{equation*}
\underbrace{2.22 \rightarrow 1.22}_{\text{\scriptsize{era}}} \rightarrow
\underbrace{4.45 \rightarrow 3.45 \rightarrow 2.45 \rightarrow 1.45}_{\text{\scriptsize{era}}} \rightarrow
\underbrace{2.22 \rightarrow 1.22}_{\text{\scriptsize{era}}} \rightarrow
\underbrace{4.45 \rightarrow 3.45 \rightarrow 2.44 \rightarrow 1.45}_{\text{\scriptsize{era}}} \rightarrow
\ldots
\end{equation*}

In the state space description of sequences (in terms of the
Mixmaster map), an \textit{epoch} is simply a point $\mathrm{P}_l$
on the Kasner circle. (It is one of the six points in the
equivalence class associated with the Kasner parameter $u_l$.)
Transitions connect epochs and thus generate the Mixmaster map.

\begin{figure}[ht]
\psfrag{a}[cc][cc]{$\Sigma_1$} \psfrag{b}[cc][cc]{$\Sigma_2$}
\psfrag{c}[cc][cc]{$\Sigma_3$} \psfrag{k}[cc][cc]{$\mathrm{T}_1$}
\psfrag{l}[cc][cc]{$\mathrm{T}_2$}
\psfrag{m}[cc][cc]{$\mathrm{T}_3$}
\psfrag{n}[cc][cc]{$\mathrm{Q}_1$}
\psfrag{o}[cc][cc]{$\mathrm{Q}_2$}
\psfrag{p}[cc][cc]{$\mathrm{Q}_3$} \centering
        \subfigure[One of the two heteroclinic cycles associated with the
          Kasner sequence generated by \mbox{$u_0 = [(1)]$}.]{
        \label{period1a}
        \includegraphics[height=0.35\textwidth]{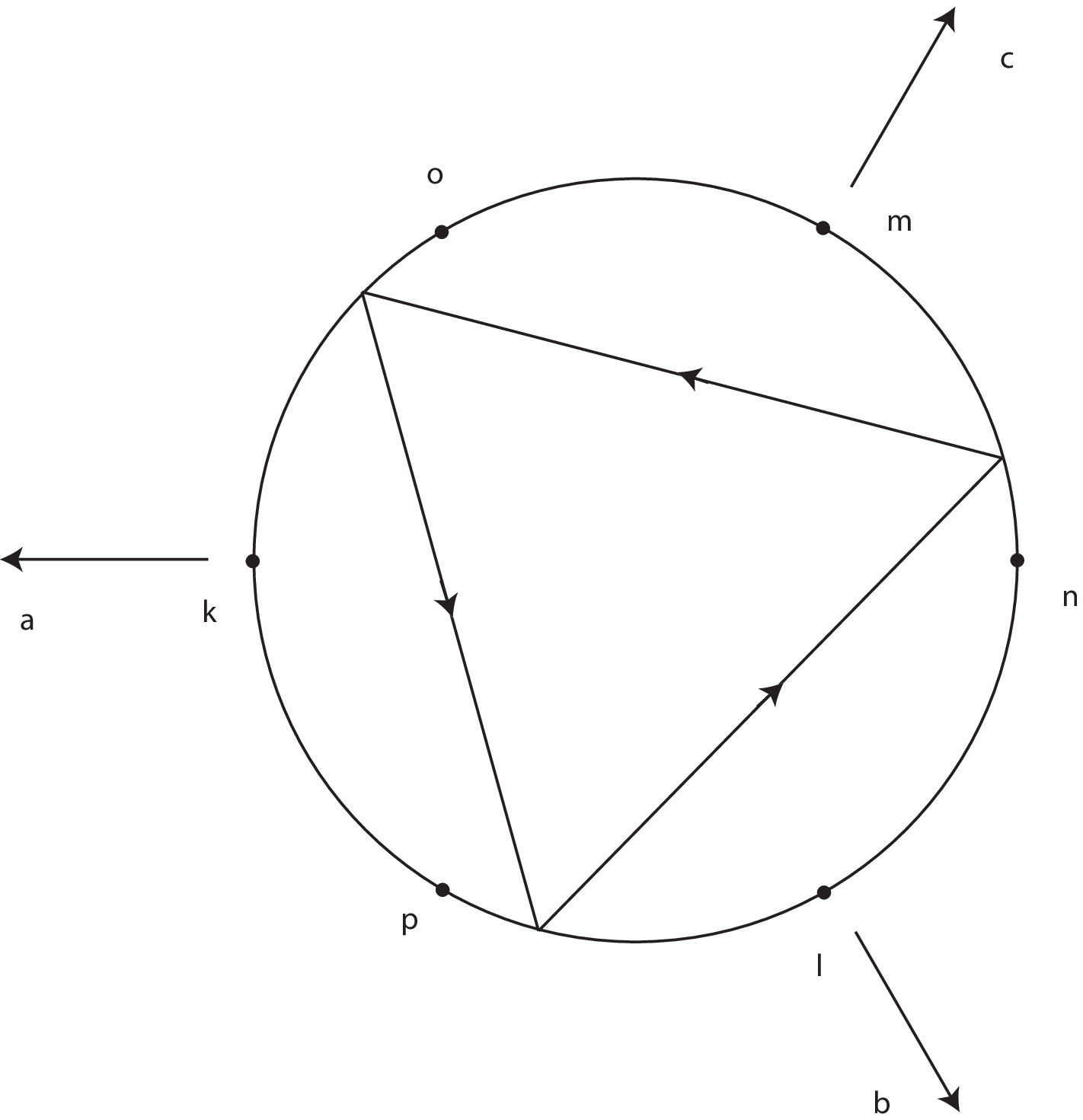}}\qquad
        \subfigure[The second of the two heteroclinic cycles associated with the
          Kasner sequence generated by \mbox{$u_0 = [(1)]$}.]{
        \label{period1b}
        \includegraphics[height=0.35\textwidth]{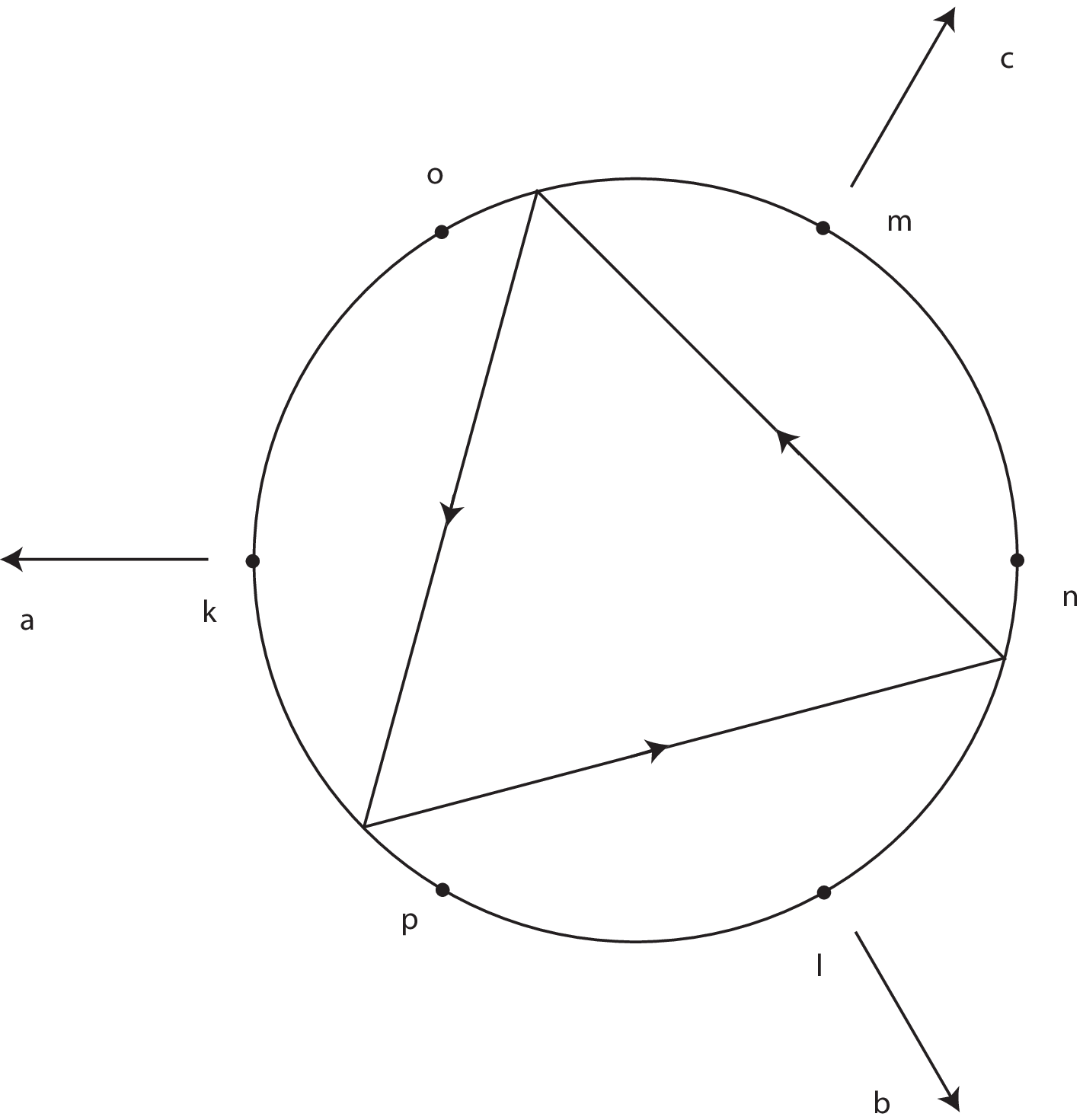}}\\
        \subfigure[One of the heteroclinic cycles associated with the Kasner
          sequence  generated by \mbox{$u_0 = [(2)]$}.]{
        \label{period2}
        \includegraphics[height=0.35\textwidth]{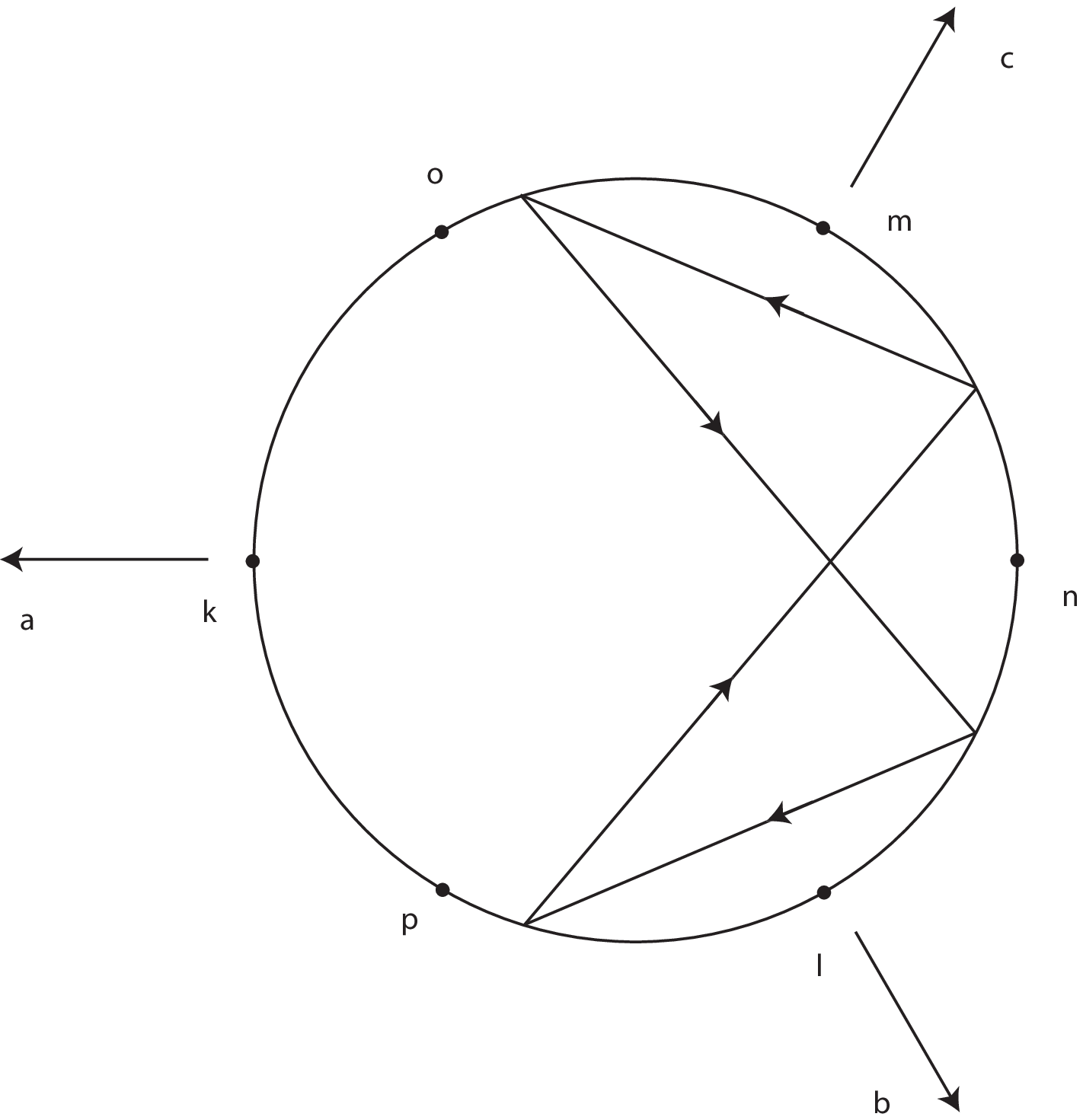}}\qquad
        \subfigure[One of the heteroclinic cycles associated with the Kasner
          sequence generated by \mbox{$u_0 = [(3)]$}.]{
        \label{period3}
        \includegraphics[height=0.35\textwidth]{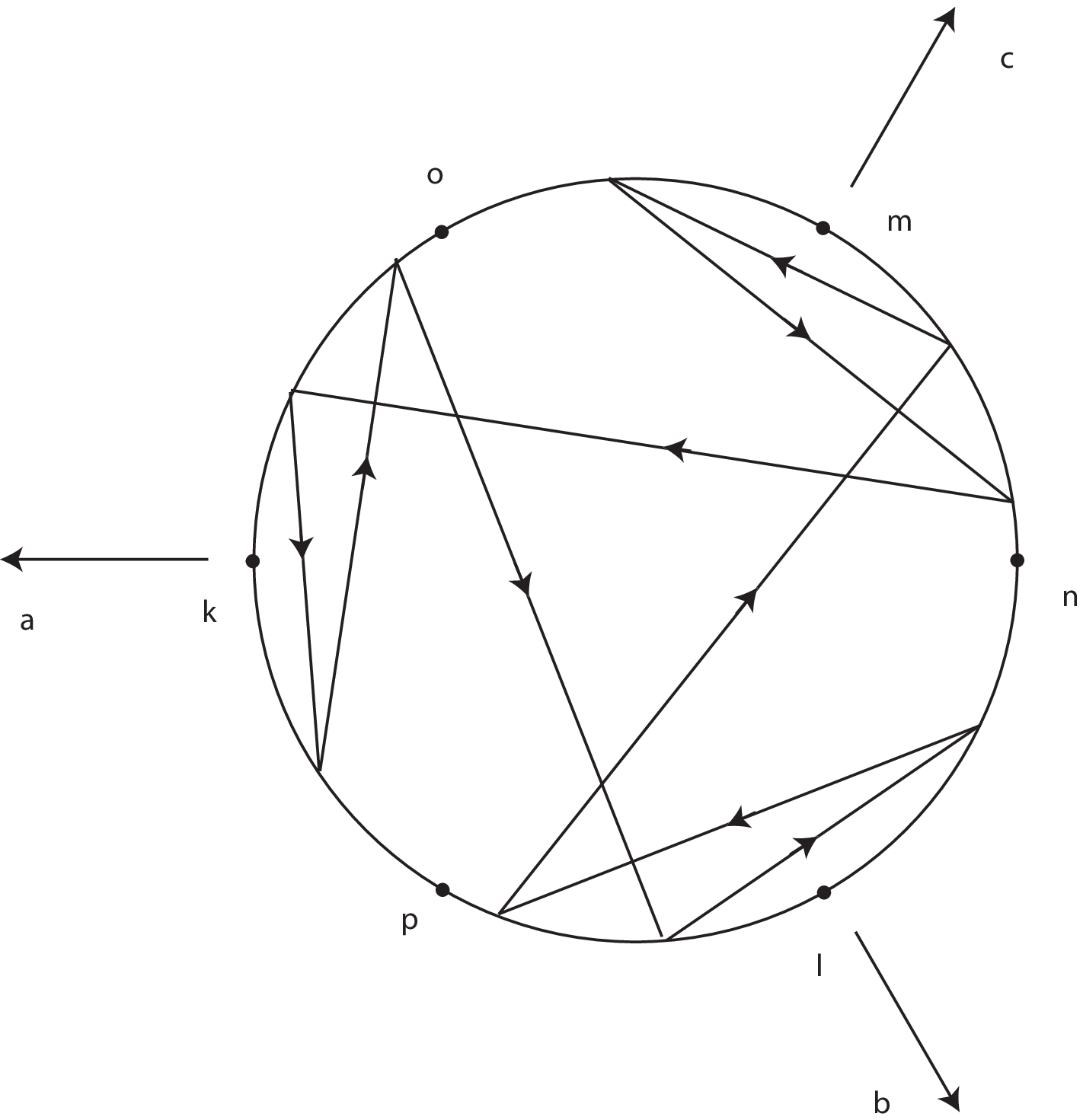}}
        \caption{Examples of heteroclinic cycles associated with
      era sequences with period $1$. The Kasner sequences have
      period $1$, $2$, and $3$, respectively; the period
      of the heteroclinic cycles is a multiple of that period.
      Note that the direction of time is towards the past.}
    \label{cycle}
\end{figure}

The Kasner parameter $u$ can be employed to measure
the (angular) distance of a point $\mathrm{P}$
on $\mathrm{K}^\ocircle$ from the Taub points or from the non-flat LRS points:
If $u \gg 1$, then $\mathrm{P}$ is at an angular distance of
approximately $u^{-1}$ from one of the Taub points.
On the other hand, in the vicinity of the non-flat LRS points (where
$u -1 \ll 1$), $u-1$ is a linear measure for the angular distance of
$\mathrm{P}$ from the closest of the non-flat LRS points.
If $u < 2$, then
$\mathrm{P}$ is closer to one of the non-flat LRS points
$\mathrm{Q}_\alpha$ than to any of the Taub points
$\mathrm{T}_\alpha$. Therefore, in the state space picture, an
\textit{era} can be described as a (finite) sequence of points
$\mathrm{K}^\ocircle \ni \mathrm{P}_l$, $\lin \leq l \leq \lout$,
obtained from the Mixmaster map, whose angular distance from the
Taub points is monotonically increasing. The era begins with a fixed
point $\mathrm{P}_{\lin}$ that is close to one of the Taub points
$\mathrm{T}_\alpha$ (the preceding point $\mathrm{P}_{\lin-1}$ was
closer to one of the other two Taub points $\mathrm{T}_\beta$,
$\beta\neq\alpha$, than to $\mathrm{T}_\alpha$); then the distance
from the Taub point $\mathrm{T}_\alpha$ monotonically increases
until, for $\mathrm{P}_{\lout}$, it exceeds the distance to the
non-flat LRS points; 
this is the terminal point for the era; it is
connected by the following transition with the initial point of the
next era; a good illustration is Figure~\ref{period3}, where each
era consists of three epochs. Note that due to
the equivalence of Kasner points, there exist several realizations
of one and the same Kasner sequence as Mixmaster sequences in the
state space picture; this is exemplified by the heteroclinic cycles
in Figure~\ref{period1a} and~\ref{period1b}.

The state space description of the cases (i)--(iv),
which are characterized by the initial Kasner parameter $\u_0 = u_0$,
is the following:
\begin{itemize}
\item[(i)] Iff $u_0 \in \mathbb{Q}$, see~\eqref{terminalki},
  the Mixmaster sequence of Kasner fixed points $(\mathrm{P}_l)_{l=0,1,\ldots}$ is finite.
  After a finite number of transitions, at the end of era $n$, the sequence
  reaches one of the LRS points $\mathrm{Q}_\alpha$ (where $u=1$); a last transition
  follows, namely the transition $\mathrm{Q}_\alpha \rightarrow \mathrm{T}_\alpha$,
  and the sequence terminates in one of the Taub points.
\item[(ii)] Iff $u_0$ is a quadratic surd, i.e., $u_0 = q_1 + \sqrt{q_2}$
  for some $q_1, q_2 \in \mathbb{Q}$ where $q_2$ is not a perfect square,
  see~\eqref{periodicki}, then
  the Mixmaster sequence of Kasner points $(\mathrm{P}_l)_{l\in\mathbb{N}}$
  is eventually periodic, where the period is a multiple of
  $(\bar{k}_1 + \cdots + \bar{k}_{\bar{n}})$; see Figure~\ref{cycle}.
  Viewed as a periodic sequence of transitions (which are heteroclinic orbits)
  we obtain a heteroclinic cycle.
  In Figure~\ref{cycle} we give some of the heteroclinic cycles
  associated with Kasner sequences with periods $1$, $2$, and $3$
  in their projection onto $(\Sigma_1,\Sigma_2,\Sigma_3)$-space.
  Note that due to permutation symmetry there are several cycles
  associated with a given periodic Kasner sequence.
\item[(iii)] Iff $u_0$ is a badly approximable irrational number, see~\eqref{boundedki},
  there exists a neighborhood of the Taub points $\mathrm{T}_\alpha$ such
  that the Mixmaster sequence of Kasner points $(\mathrm{P}_l)_{l\in\mathbb{N}}$ does not
  enter this neighborhood. This is simply because there exists a maximal
  value of the sequence $(u_l)_{l\in\mathbb{N}}$.
\item[(iv)] Iff $u_0$ is a well approximable irrational number, see~\eqref{wellki},
  then the Mixmaster sequence $(\mathrm{P}_l)_{l\in\mathbb{N}}$ comes arbitrarily close
  to the Taub points.
  This is the generic case.
\end{itemize}

In the following we analyze the generic case (iv) in more detail.
Let $u_0$ be a well approximable irrational number, i.e., a number whose continued
fraction expansion
\begin{equation}\label{u0cont}
  u_0 = \big[ k_0; k_1, k_2 , k_3, \dotsc \big]
\end{equation}
defines an unbounded sequence $(k_i)_{i\in\mathbb{N}}$. By
construction, era number $i$ contains $k_i$ epochs (which we call
its length). A natural question to ask concerns the distribution of
the partial quotients $k_i$. For a `typical' well approximable irrational
number, how often does the number $1$ appear in the sequence
$(k_i)_{i\in\mathbb{N}}$? How often the number $2$? And what about
the number $1000$? The answer is given by Khinchin's
law~\cite{kin64}.
Let $P_n(k{=}m)$ denote the probability that a randomly chosen
partial quotient among $(k_1,\dotsc,k_n)$ equals $m\in\mathbb{N}$. In the
asymptotic limit, i.e., for $P(k{=}m) =
\lim_{n\rightarrow\infty} P_n(k{=}m)$ we have
\begin{equation}\label{Khinchin}
P(k{=}m)   \,= \,  {}^2\!\log \big( \frac{m+1}{m+2}\big) -{}^2\!\log \big( \frac{m}{m+1} \big) \:,
\end{equation}
i.e., the partial quotients of the continued fraction representation of~\eqref{u0cont} are
distributed like a random variable whose probability distribution is given by~\eqref{Khinchin}.
(By~\eqref{Khinchin}, the number $1$ appears in $42\%$ of the slots, the number $2$ in $17\%$, and
the number $1000$ in $1.4 * 10^{-6}\%$ of the slots.)
Khinchin's law applies for \emph{almost all} numbers $u_0$.

For a (generic) Kasner sequence $(u_l)_{l\in\mathbb{N}}$ with initial parameter
$u_0 = [k_0; k_1, k_2,\ldots]$
and its associated era sequence $(\u_s)_{s\in\mathbb{N}}$,
where $\u_s = [k_s; k_{s+1}, k_{s+2},\ldots]$,
the expression $P(k{=}m)$ of~\eqref{Khinchin}
represents the probability that
a randomly chosen era of $(\u_s)_{s\in\mathbb{N}}$ has length $m$;
this corresponds to the probability that
the initial value of an era is contained in the interval $[m,m+1)$.
In this manner, the probability distribution~\eqref{Khinchin}
makes possible a stochastic interpretation of generic Kasner sequences.

The probability distribution~\eqref{Khinchin} results in
extraordinary properties of the Mixmaster/Kasner map,
which will be of crucial importance in the considerations
of Section~\ref{beliefs}.
We do not discuss details here but
refer to future work; however, we cannot refrain from
giving a teaser:
For a generic Kasner sequence $(u_l)_{l\in\mathbb{N}}$
and its associated era sequence $(\u_s)_{s\in\mathbb{N}}$ there exist
infinitely many eras such that the length (i.e., the number of epochs)
of the $n$\raisebox{0.7ex}{\small th} era is larger than $n \log n$;
however, for sufficiently large $n$, the length is guaranteed
to be bounded by $n \log^2 n$. (For a proof of this result
by Borel and Bernstein see~\cite{Hardy/Wright:1979}; see
also~\cite{Flajolet/Vallee/Vardi}.)
Properties of this kind underline 
the remarkable intricacies
of the heteroclinic structures on the Mixmaster attractor.

\section{Mixmaster facts}
\label{furthermix}

In this section we turn to what is known about the past
asymptotic dynamics of generic type~IX solutions.
The main Mixmaster fact is Ringstr\"om's `Bianchi type~IX attractor theorem'.

\subsection*{The main Mixmaster fact} \label{main}

Consider a solution of Bianchi type~IX that is either vacuum or
associated with a perfect fluid satisfying $-\textfrac{1}{3}< w
< 1$. Recall that such a solution is called \textit{generic} if
it is not past asymptotically self-similar, i.e., if its
$\alpha$-limit set is neither the point $\mathrm{F}$, nor any
of the points $\mathrm{CS}_\alpha$, nor a point on
$\mathrm{TL}_\alpha$; in other words, a generic solution
corresponds to an orbit in $\BIX$ that is neither contained in
$\mathscr{F}$, nor in $\mathcal{C\!S}_\alpha$, nor in
$\mathcal{L\!R\!S}_\alpha$. Therefore, the set of generic
Bianchi type~IX states is an open set in $\BIX$. (To conform
with~\cite{waiell97,rin01} we use the future directed time
variable $\tau$.)

The main results concerning generic Bianchi type~IX models are
due to Ringstr\"om~\cite{rin01}; these results rest on earlier
work that is reviewed and derived in~\cite{waiell97}, and
on~\cite{ren97,rin00}. In the following we state the main
theorem in a version adapted to our purposes.

\begin{theorem}[\cite{rin01}]\label{rinthm}
Let $(\Sigma_1,\Sigma_2,\Sigma_3,N_1,N_2,N_3)(\tau)$ be a
generic solution of Bianchi type~IX, i.e., a generic solution
of~\eqref{IXeq} in $\BIX$. Then
\begin{equation}\label{rinthmeq}
\Delta_{\mathrm{II}} =  N_1 N_2 + N_2 N_3 + N_3 N_1 \:\rightarrow \:0
\qquad\text{and}\qquad
\Omega \,\rightarrow\, 0
\end{equation}
as $\tau\rightarrow-\infty$.
\end{theorem}
Note that this Theorem applies to both the fluid and the vacuum
case; in the latter case~\eqref{rinthmeq} becomes
$\Delta_{\mathrm{II}} \rightarrow 0$ and $\Omega\equiv 0$.

The proof of Theorem~\ref{rinthm} given in~\cite{rin01} is
delicate. In the first part it is proved that the
$\alpha$-limit set of a generic Bianchi type~IX solution is
non-empty and must contain a point on the Kasner circle. The
main part of the proof deals with the fact that the function
$\Delta_{\mathrm{II}}(\tau)$ is in general not monotone. There
exist times where $\Delta_{\mathrm{II}}(\tau)$ increases (as
$\tau\rightarrow -\infty$); the associated growth must
therefore be controlled and shown to be negligible compared to
the overall decrease in $\Delta_{\mathrm{II}}$. This is done by
a careful analysis of the equations and (approximate)
solutions. In~\cite{heiuggproof} we give an alternative and
relatively short and succinct proof of Theorem~\ref{rinthm}
which is based on an in-depth understanding of the hierarchical
structure of the dynamical system~\eqref{IXeq} (as represented
by Figure~\ref{contraction}). It is important to note, however,
that either of the proofs fail in the other Bianchi types that
are conjectured to exhibit an oscillatory approach towards the
singularity, i.e., the proofs fail for
types~$\mathrm{VI}_{-1/9}$ and VIII. This is unfortunate, since
there are reasons to believe that these models are more
relevant than type~IX as regards the dynamics of generic
(inhomogeneous) cosmologies, see~\cite{heiuggproof}.

Using the concept of the Mixmaster attractor, cf.~\eqref{AIXdef},
we obtain an
equivalent formulation of Theorem~\ref{rinthm}: Let $X(\tau) =
(\Sigma_1,\Sigma_2,\Sigma_3,N_1,N_2,N_3)(\tau)$ be a generic
solution of Bianchi type~IX. Then
\begin{equation}\label{distfromatt}
\| X(\tau) - \mathcal{A}_{\mathrm{IX}} \| \rightarrow 0 \qquad
(\tau\rightarrow -\infty)\:,
\end{equation}
where the distance $\|X -  \mathcal{A}_{\mathrm{IX}}\|$ is
given as $\min_{Y\in\mathcal{A}_{\mathrm{IX}}} \| X- Y\|$.

Theorem~\ref{rinthm} thus states that the attractor of generic
type~IX solutions resides on $\mathcal{A}_{\mathrm{IX}}$;
however, whether the past attractor is in fact
$\mathcal{A}_{\mathrm{IX}}$ or merely a subset thereof remains
open. (The terminology `Mixmaster attractor' is seductive but
might turn out to be quite misleading.) Likewise, the theorem
does not provide any direct information about the details of the
asymptotic behavior of solutions.

\subsection*{Consequences}

The prerequisite for a deeper
understanding of the asymptotic behavior and the oscillatory
dynamics of generic type~IX solutions is an
understanding of the Mixmaster attractor.
In Section~\ref{maps} we have identified the structures on the
Mixmaster attractor $\mathcal{A}_{\mathrm{IX}}$ that are
induced by the flow of the dynamical system: heteroclinic
cycles [case (ii)] and finite [case (i)] and infinite
heteroclinic sequences [cases (iii) and (iv)]. All these
structures qualify (a priori) as possible $\alpha$-limit sets
of generic type~IX orbits. In conjunction with these results, the main
theorem~\ref{rinthm}
implies a number of  further facts about the
attractor---`Mixmaster attractor facts', which we give as a
list of \textbf{corollaries}. (For proofs
see~\cite{heiuggproof}, and also~\cite{rin01}.)

\begin{enumerate}
\item\label{possessesalpha} A generic, i.e. not past asymptotically
    self-similar, type~IX orbit possesses an $\alpha$-limit
    point on $\mathcal{A}_{\mathrm{IX}}$.
\item\label{onethenall} If $\mathrm{P}\in \mathcal{A}_{\mathrm{IX}}$ is
    an $\alpha$-limit point of a type~IX orbit, then the
    entire heteroclinic cycle/sequence (Mixmaster sequence)
    through $\mathrm{P}$ must be contained in the
    $\alpha$-limit set.
\item\label{taubthenmany} If one of the Taub points
    $\{\mathrm{T}_1, \mathrm{T}_2, \mathrm{T}_3\}$ is an
    $\alpha$-limit point of a type~IX orbit, then the
    $\alpha$-limit set contains Kasner fixed points
    associated with arbitrarily large values of the Kasner
    parameter $u$.
\item For generic solutions of Bianchi type~IX the Weyl
    curvature scalar $C_{abcd}C^{abcd}$ (and therefore also
    the Kretschmann scalar) becomes unbounded towards the
    past.
\item Taking into account both the expanding and
    contracting phases of Bianchi type~IX solutions,
    generic Bianchi type~IX initial data generate an
    inextendible maximally globally hyperbolic development
    associated with past and future singularities where the
    curvature becomes unbounded.
\item\label{uniformcor}
  Convergence to the Mixmaster attractor is uniform
    on compact sets of generic initial data: Let
    $\mathscr{X}$ be a compact set in $\BIX$ that does not
    intersect any of the manifolds $\mathscr{F}$,
    $\mathcal{C\!S}_\alpha$, $\mathcal{L\!R\!S}_\alpha$, so
    that each initial data $\mathring{x} \in \mathscr{X}$
    generates a generic
    type~IX solution. 
    Let $X(\mathring{x};\tau)$ denote the type~IX solution with
    $X(\mathring{x},0) = \mathring{x}$. Then
    \begin{equation}
      \| X(\mathring{x}; \tau) - \mathcal{A}_{\mathrm{IX}} \| \rightarrow 0 \qquad (\tau\rightarrow -\infty)
    \end{equation}
    uniformly in $\mathring{x} \in \mathscr{X}$.
\end{enumerate}

Corollary 3 implies that the $\alpha$-limit set contains an
infinite set of Kasner fixed points in a neighborhood of the
Taub point(s), but this set is not necessarily a continuum of
fixed points; cf.~the previous discussion about possible
$\alpha$-limit sets. If a Kasner point with $u\in\mathbb{Q}$ is
contained in the $\alpha$-limit of an orbit, so is a Taub
point. This is an immediate consequences of the results of
Section~\ref{maps}, case (i). On the other hand, if the
$\alpha$-limit set of an orbit is a heteroclinic cycle or a
heteroclinic sequence (possibly in combination with a cycle)
associated with cases (ii) and (iii) of Section~\ref{maps},
then there exists a neighborhood of the Taub points whose
intersection with the $\alpha$-limit set is empty.

Note that the oscillatory behavior of asymptotic type~IX
dynamics, which we unfortunately know no details about,
constitutes an example of asymptotic self-similarity
breaking~\cite{limetal06}. In order to make progress as regards
the details of the asymptotic oscillatory behavior, it is
natural to first establish the connection between the
Mixmaster/Kasner map and dynamics for a finite time interval
$\Delta\tau$, discussed next.

\subsection*{Finite Mixmaster shadowing}

To make contact between the Mixmaster map and Bianchi type~IX
asymptotic dynamics we introduce the concept of \textit{finite
Mixmaster shadowing\/} which formalizes the following basic
idea: Given a sequence of transitions we can choose type~IX
initial data sufficiently close to the initial data of the
sequence so that the type~IX solution generated by this data
remains close to the sequence for some `time'.

Let $\mathrm{P}_0$ be a Kasner fixed point (but $\mathrm{P}_0
\not\in \{\mathrm{T}_1, \mathrm{T}_2,\mathrm{T}_3\}$) and let
$u_0$ be the associated value of the Kasner parameter. There
exists a unique sequence of transitions
$(\mathcal{T}_l)_{l\in\mathbb{N}}$ and an associated Mixmaster
sequence $(\mathrm{P}_l)_{l \in\mathbb{N}}$ with $\mathrm{P}_0$
as initial data; the associated Kasner sequence is
$(u_l)_{l\in\mathbb{N}}$. (If $u_0 \in \mathbb{Q}$, the
sequence terminates at one of the Taub points after a finite
number of transitions.) We shall make the definition that a
type~IX solution shadows a finite piece $(\mathcal{T}_l)_{l
=0,1,\ldots, L}$ of the sequence of transitions if it is
contained in a prescribed (small) tubular neighborhood of
$(\mathcal{T}_l)_{l =0,1,\ldots, L}$. However, a standard
$\epsilon$-neighborhood of the sequence fails to be a
reasonable measure of closeness, because in the vicinity of a
Taub point the transitions lie so dense that consecutive
transitions are not separated from each other by their
respective $\epsilon$-neighborhoods. Therefore, the
introduction of \textit{adapted} tubular neighborhoods is
necessary to take into account the sensitivity of the flow at
the Taub points and to capture more accurately the intuitive
idea of shadowing.

Let $\epsilon > 0$ be small. A `\textit{Taub-adapted
neighborhood}' of the Mixmaster sequence $(\mathrm{P}_l)_{l
\in\mathbb{N}}$ is a sequence $(\mathcal{U}_l)_{l
\in\mathbb{N}}$ of open balls, where $\mathcal{U}_l$ is
centered at the point $\mathrm{P}_l$ and has radius $\epsilon
u_l^{-2}$, i.e., $\mathcal{U}_l = \{ X \in \overlineBIX\,:\, \|
X - \mathrm{P}_l\| < \epsilon u_l^{-2} \}$. (The radius
$\epsilon u_l^{-2}$ is chosen to ensure that the intersection
of the ball $\mathcal{U}_l$ with the Kasner circle induces more
or less a standard $\epsilon$-neighborhood $(u_l - \epsilon,
u_l + \epsilon)$ of the Kasner parameter $u_l$; recall from
Section~\ref{maps} that $u^{-1}$ measures the angular distance
of a fixed point from a Taub point.) A Taub-adapted tubular
neighborhood of the sequence of transitions
$(\mathcal{T}_l)_{l\in\mathbb{N}}$ is the sequence of tubes
$(\mathcal{V}_l)_{l\in\mathbb{N}}$ that linearly interpolate
between $\mathcal{U}_l$ and $\mathcal{U}_{l+1}$. Based on this
definition we say that a type~IX solution $X(\tau)$
\textit{shadows} a finite piece $(\mathcal{T}_l)_{l
=0,1,\ldots, L}$ of the sequence of transitions if it moves in
a prescribed Taub-adapted tubular neighborhood
$(\mathcal{T}_l)_{l =0,1,\ldots, L}$, i.e., if there exists a
sequence of times $(\tau_l)_{l= 0,1,\ldots,L+1}$ such that
$X(\tau)$ is contained in $\mathcal{V}_l$ for all $\tau \in
(\tau_{l+1},\tau_l]$ for all $0\leq l \leq L$.

Making use of these concepts, a formulation of finite Mixmaster
shadowing is the following: Let $\epsilon > 0$ and $L \in
\mathbb{N}$. Consider the sequence of transitions
$(\mathcal{T}_l)_{l \in \mathbb{N}}$ emanating from an initial
Kasner point $\mathrm{P}_0$ and its Taub-adapted tubular
neighborhood (associated with $\epsilon$). Then there exists
$\delta_\epsilon >0$ such that each type~IX orbit $X(\tau)$
that is generated by initial data $X_0$ with $\|X_0 -
\mathrm{P}_0\| <\delta_\epsilon$ shadows the finite piece
$(\mathcal{T}_l)_{l =0,1,\ldots, L}$ of the sequence of
transitions. (A proof of this statement---in a slightly
different form---has been given by Rendall~\cite{ren97}.
Alternatively, one can invoke the regularity of the dynamical
system, the center manifold reduction theorem and continuous
dependence on initial data.) Evidently, $\delta_\epsilon$
depends on the choice of $\epsilon$ and $L$. More importantly,
however, $\delta_\epsilon$ depends on the position of
$\mathrm{P}_0$---shadowing is not uniform; in particular, if we
consider a series of initial points that approach one of the
Taub points, then $\delta_\epsilon$ necessarily converges to
zero along this series---shadowing is more delicate in the
vicinity of a Taub point. We will return to this issue in some
detail in the next Section.

Finite Mixmaster shadowing concerns any generic type~IX orbit
$X(\tau)$. Let $\mathrm{P} \in \mathrm{K}^\ocircle$ be an
$\alpha$-limit point of the type~IX orbit $X(\tau)$; without
loss of generality we may assume that $\mathrm{P}$ is not one
of the Taub points. (The existence of such a point is ensured
by Corollaries~\ref{possessesalpha}--\ref{taubthenmany} of Section~\ref{furthermix}.)
For
simplicity we assume that $\mathrm{P}$ is associated with an
irrational value of the Kasner parameter, which guarantees that
the sequence $(\mathrm{P}_l)_{l\in\mathbb{N}}$ emanating from
$\mathrm{P} = \mathrm{P}_0$ is an infinite sequence. Since
$\mathrm{P}$ is an $\alpha$-limit point of $X(\tau)$, there
exists a sequence of times $(\tau_n)_{n\in\mathbb{N}}$, $\tau_n
\rightarrow -\infty$ ($n\rightarrow \infty$), such that
$X(\tau_n) \rightarrow \mathrm{P}$ ($n\rightarrow \infty$).
Therefore, we observe a recurrence of phases, where the orbit
$X(\tau)$ shadows $(\mathrm{P}_l)_{l\in\mathbb{N}}$ with an
increasing degree of accuracy, i.e., shadowing takes place for
increasingly longer pieces of the sequence or in ever smaller
neighborhoods.
(If the
Kasner parameter of $\mathrm{P}$ is rational, then the sequence
$(\mathrm{P}_l)_{l\in\mathbb{N}}$ is finite and terminates at a
Taub point. $X(\tau)$ will shadow this finite sequence
recurrently with an increasing degree of precision.)
We will return to this issue in some more detail in
the subsection `Stochastic Mixmaster beliefs' of Section~\ref{beliefs}.

The concept of shadowing leads directly to the concept of
\textit{approximate sequences} which we introduce next.
Consider a generic type~IX orbit $X(\tau) =
(\Sigma_1,\Sigma_2,\Sigma_3,N_1,N_2,N_3)(\tau)$ and the
function $\|X(\tau) - \mathrm{K}^\ocircle\|$, where the
distance $\|X - \mathrm{K}^\ocircle\|$ is given as $\min_{Y\in
\mathrm{K}^\ocircle} \| X- Y\|$. When the orbit $X(\tau)$
traverses a (sufficiently small) neighborhood of a fixed point
on $\mathrm{K}^\ocircle\backslash \{\mathrm{T}_1, \mathrm{T}_2,
\mathrm{T}_3\}$, the function $\|X(\tau) -
\mathrm{K}^\ocircle\|$ exhibits a unique local minimum. This is
immediate from the transversal hyperbolic saddle structure of
the fixed point. (However, the flow in the vicinity of the Taub
points is more intricate, since these points are not
transversally hyperbolic.) It follows that the function
$\|X(\tau) - \mathrm{K}^\ocircle\|$ can be used to partition
$X(\tau)$ into a sequence of segments
in a straightforward manner:%
\footnote{The definition of a partition of $X(\tau)$ into
  segments
  seems natural; it is important to note, however,
  that any definition depends on the formulation of the
  problem
  and is to a certain extent arbitrary.
  For instance, instead of using the minima of
  $\|X(\tau) - \mathrm{K}^\ocircle\|$ one might prefer
  to analyze the projection of the orbit
  onto $(\Sigma_1,\Sigma_2,\Sigma_3)$-space and use
  the extrema of $\Sigma^2(\tau)$.
  However, the conclusions drawn from
  any construction of segments are quite insensitive
  to the details of the definition.}
The local minima of $\|X(\tau) - \mathrm{K}^\ocircle\|$ form an
infinite sequence $(\tau_l)_{l\in\mathbb{N}}$ such that $\tau_l
\rightarrow -\infty$ as $l\rightarrow \infty$. (This follows
directly from Corollaries~\ref{possessesalpha}
and~\ref{onethenall} because $X(\tau)$ has $\alpha$-limit
point(s) on the Kasner circle and $\alpha$-limit points on the
type~II subset.) A \textit{segment} of $X(\tau)$ is defined to
be the solution curve between two consecutive minima, i.e., the
image of the interval $(\tau_{l+1}, \tau_l]$. (Note that $\tau
\rightarrow -\infty$ in the approach to the singularity, while
the discrete `time' $l$ is past-directed, i.e., $l\rightarrow
\infty$ towards the singularity. This convention is chosen to
agree with the standard convention for the Mixmaster and the
Kasner map.) In the asymptotic regime, i.e., in the approach to
the Mixmaster attractor, finite shadowing entails that a finite
sequence of segments will resemble a finite sequence of type~II
transitions; this assumes, however, that the type~IX orbit does
not come too close to any of the Taub points. (In the
neighborhood of a Taub point the flow of the dynamical system
is much more intricate. This might yield recurring
interruptions of the `standard behavior'.) We call the type~IX
orbit in its segmented form an \textit{approximate sequence of
transitions}.

\begin{figure}[ht]
\psfrag{Xl}[rc][rc][0.9][0]{$X(\tau_l)$}
\psfrag{Xl1}[lc][lc][0.9][0]{$X(\tau_{l+1})$}
\psfrag{Pl}[lc][lc][0.9][0]{$\checkP_l$}
\psfrag{Pl1}[rc][rc][0.9][0]{$\checkP_{l+1}$}
\psfrag{vl}[cc][cc][0.7][0]{$\check{\delta}_l$}
\psfrag{X}[cc][cc][1][0]{$X(\tau)$}
\psfrag{K}[cc][cc][1][0]{$\mathrm{K}^\ocircle$}
\psfrag{mm}[cc][cc][0.5][2]{Mixmaster map}
\psfrag{km}[tc][tc][0.5][2]{Kasner map}
\psfrag{dP}[rc][rc][0.9][0]{$(\norDelta\check{u}_l \leftrightarrow)\, \norDelta\checkP_l$}
\centering
\includegraphics[height=0.35\textwidth]{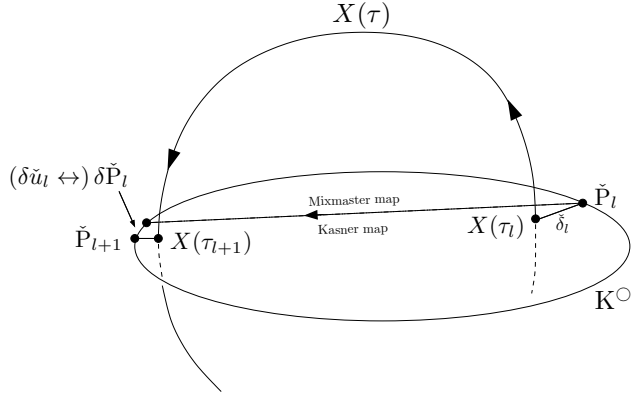}
\caption{A type~IX orbit decomposes into segments $X(\tau_l)\rightarrow X(\tau_{l+1})$.
  The points $X(\tau_l)$, $l \in\mathbb{N}$, are the local minima of
  the distance between $X(\tau)$ and $\mathrm{K}^\ocircle$.
  The `check point' $\checkP_l$ is the point on $\mathrm{K}^\ocircle$ that is closest
  to $X(\tau)$ (for $\tau$ in a neighborhood of $\tau_l$).
  Since each check point $\checkP_l$ is associated with a value $\check{u}_l$
  of the Kasner parameter, the `approximate Mixmaster sequence' $(\checkP_l)_{l\in\mathbb{N}}$ induces
  an `approximate Kasner sequence' $(\check{u}_l)_{l\in\mathbb{N}}$.
  In general, $\checkP_{l+1}$/$\check{u}_{l+1}$ are
  not generated from $\checkP_l$/$\check{u}_l$ by the exact
  Mixmaster/Kasner map~\eqref{Kasnermap}, but differ by an
  error $\norDelta\checkP_l$/$\norDelta \check{u}_l$.}
\label{checkmap}
\end{figure}

In addition, we define a sequence of `check points' that is
associated with an approximate sequence of transitions. Each
minimum $\tau_l$ of $\|X(\tau) - \mathrm{K}^\ocircle\|$ is
associated with a Kasner fixed point $\check{\mathrm{P}}_l$ (a
`check point') that is defined as the minimizer on
$\mathrm{K}^\ocircle$ of the distance between $X(\tau_l)$ and
$\mathrm{K}^\ocircle$. (Note that the check points
$(\check{\mathrm{P}}_l)_{l\in\mathbb{N}}$ do not lie on the
type~IX orbit $X(\tau)$.) Since each check point
$\check{\mathrm{P}}_l$ is associated with a value $\check{u}_l$
of the Kasner parameter, the sequence
$(\check{\mathrm{P}}_l)_{l\in\mathbb{N}}$ induces a sequence
$(\check{u}_l)_{l\in\mathbb{N}}$. The value $\check{u}_{l+1}$
is in general not generated from $\check{u}_l$ by the exact
Kasner map~\eqref{Kasnermap}, but differs from that value by an
error of $\norDelta \check{u}_l$. In our terminology, the
sequence $(\check{\mathrm{P}}_l)_{l\in\mathbb{N}}$ is an
approximate Mixmaster sequence;
$(\check{u}_l)_{l\in\mathbb{N}}$ is an approximate Kasner
sequence, see Figure~\ref{checkmap}.

The approximate Kasner sequence $\check{u}_l$ associated with a
type~IX orbit $X(\tau)$ does not follow the Kasner
map~\eqref{Kasnermap} exactly. A natural question to ask,
however, is whether the errors $\norDelta \check{u}_l$ of the
approximate Kasner sequence converge to zero as $l\rightarrow
\infty$ or not. If $\norDelta \check{u}_l \rightarrow 0$ as
$l\rightarrow \infty$ for a type~IX orbit, this means that its
dynamics is completely described by an `asymptotic
Mixmaster/Kasner map', i.e., by a map that converges to the
Mixmaster/Kasner map towards the singularity. However, if
$\norDelta \check{u}_l \not\rightarrow 0$ as $l\rightarrow
\infty$, then the evolution is interrupted
repeatedly---infinitely many times---by phases where the
dynamics is completely different from the Mixmaster dynamics,
e.g., `eras of small oscillations' \cite{bkl70}. (In the
present state space description an era of small oscillations is
associated with type~$\mathrm{VII}_0$ behavior in the vicinity
of one of the Taub lines $\mathrm{TL}_\alpha$, where $N_\beta$
and $N_\gamma$ are small and of the same order; for details we
refer to the discussion of type~$\mathrm{VII}_0$ dynamics
in~\cite{heiuggproof}.) In the subsection `Stochastic Mixmaster
beliefs' of Section~\ref{beliefs} we will investigate the behavior of
$\norDelta\check{u}_l$ along type~IX orbits in detail.

\section{Mixmaster beliefs}
\label{beliefs}

\subsection*{Attractor beliefs}

There remain several important open problems. In the following
we will address the most pressing questions; in particular we
will transform vague beliefs into refutable conjectures, and
give arguments in their favor.

An immediate question is the following:
What are the actual $\alpha$-limit sets of type~IX orbits?
Consider a type~IX orbit $\gamma$. The $\alpha$-limit set $\alpha(\gamma)$
contains a number of Kasner fixed points (and the associated type~II transitions),
each of which is characterized by a particular value of the Kasner parameter $u$.
The set of Kasner parameters obtained in this way from $\alpha(\gamma)$
we denote by $U(\gamma)$. The question of which form $U(\gamma)$
can take for different orbits $\gamma$ is open, the
most significant issues being the following:

\begin{itemize}
\item[(i)] Is it possible that there exists $\gamma$ whose limit set
  $U(\gamma)$ consists of rational numbers only?
  It is straightforward to exclude that $U(\gamma)$ coincides with $\mathbb{Q}$
  itself (or a dense subset thereof), the reason being that $\alpha$-limit sets
  are necessarily closed. Another a priori constraint is that
  $U(\gamma)$ must contain $u = \infty$ (which
  characterizes the Taub points) if
  $U(\gamma)$ contains a rational number; see Section~\ref{maps}.
  Corollary~\ref{taubthenmany} of Section~\ref{furthermix}
  then implies that $U(\gamma)$ is unbounded,
  i.e., $U(\gamma)$ contains
  arbitrarily large values of the Kasner parameter.
  A set that is compatible with these basic requirements is, e.g.,
  $U(\gamma) = \mathbb{N} \cup \{\infty\}$. Whether there exist
  orbits $\gamma$ such that $U(\gamma)$ takes this (or a related) form
  remains open.
%
  %
\item[(ii)] Is it conceivable that there exist orbits $\gamma$
 such that $\alpha(\gamma)$ is a heteroclinic cycle, see Figure~\ref{cycle}?
 In this case the set $U(\gamma)$ is generated by a quadratic surd
 $u = [(\bar{k}_1, \bar{k}_2,\ldots,\bar{k}_n)]$ via the Kasner map;
 see Section~\ref{maps}; in particular, $U(\gamma)$ is finite.
 However, whether orbits $\gamma$ with this particular past asymptotic
 behavior really exist is an open problem.
\item[(iii)] Can there exist orbits $\gamma$ such that
    $U(\gamma)$ is bounded but contains infinitely many
    $u$-values? Is the Kasner sequence generated by a badly
    approximable number a candidate? The Kasner sequence
    $(u_l)_{l\in\mathbb{N}}$ generated by a badly
    approximable number is an infinite sequence that is
    bounded. However, there must be at least one
    accumulation point of this sequence; if $U(\gamma)$
    contains $(u_l)_{l\in\mathbb{N}}$, then $U(\gamma)$
    must contain the accumulation point of the sequence as
    well (since $\alpha$-limit sets are closed). If this
    accumulation point is a well approximable number,
    $U(\gamma)$ cannot be bounded; however, if the
    accumulation point is a quadratic surd, no
    inconsistencies arise. Hence, a priori, there might
    exist orbits $\gamma$ such that $U(\gamma)$ is not
    finite but still bounded. Whether this is indeed the
    case is doubtful, but hard to exclude a priori.
\end{itemize}

An open problem that might be quite separate from the questions raised above
concerns the behavior of all type~IX orbits
save a set of measure zero.

\begin{definition}
The past attractor
of a dynamical system given on a state space $X$
is defined as the smallest closed
invariant set $\mathcal{A}^- \subseteq \overline{X}$ such that
$\alpha(p) \subseteq \mathcal{A}^-$
for all $p \in X$ apart from a set of measure zero \cite{mil85}.
\end{definition}

\begin{conjecture}[Mixmaster attractor conjecture]
The past attractor of the type~IX dynamical system
coincides with the Mixmaster attractor
$\mathcal{A}_{\mathrm{IX}}$
(rather than being a subset thereof).
\end{conjecture}

Why is this a belief and not a fact? Theorem~\ref{rinthm}
implies that $\mathcal{A}^- \subseteq
\mathcal{A}_{\mathrm{IX}}$; however, it is believed that
$\mathcal{A}^- = \mathcal{A}_{\mathrm{IX}} =
\mathrm{K}^\ocircle \cup \mathcal{B}_{N_1}^{\mathrm{vac.}} \cup
\mathcal{B}_{N_2}^{\mathrm{vac.}} \cup
\mathcal{B}_{N_3}^{\mathrm{vac.}}$. (The usage of the
terminology `Mixmaster attractor' for the set
$\mathcal{A}_{\mathrm{IX}}$ reflects the strong belief in the
Mixmaster conjecture.) It is difficult to imagine how the
Mixmaster attractor conjecture could possibly be violated. For
instance, it seems rather absurd that the past attractor
consists only of (a subset of) heteroclinic cycles---but there
exist no proofs. A closely related belief is the following
stronger statement.

\begin{conjecture}
For almost all Bianchi type~IX solutions $\gamma$ the $\alpha$-limit set
$\alpha(\gamma)$ coincides with the Mixmaster attractor $\mathcal{A}_{\mathrm{IX}}$.
\end{conjecture}

We use the term `almost all' in a noncommittal way without
specifying the measure; recall that the word `generic' already
has the well-defined meaning of `not past asymptotically
self-similar'. (The usage of the word `generic' is in accord
with~\cite{rin01}.)

\subsection*{Stochastic beliefs}\label{stochasticbeliefs}

This subsection is concerned with the (open) question of which
role the Mixmaster/Kasner map 
plays in the asymptotic evolution of type~IX solutions. The
basis for our considerations are the results of
Section~\ref{maps} where we 
discussed the Mixmaster/Kasner map and the stochastic aspects
of (generic) Kasner sequences. The \textit{Mixmaster
stochasticity conjecture} supposes that these stochastic
properties carry over to almost every type~IX orbit when
represented as an approximate Kasner sequence.

\begin{conjecture}[Mixmaster stochasticity]
The approximate Kasner sequence
$(\check{u}_l)_{l\in\mathbb{N}}$ associated with a generic
type~IX orbit admits a stochastic interpretation in terms of
the probability distribution associated with the Kasner map,
cf.~Section~\ref{maps}. (This holds with probability one, i.e.,
for almost every generic type~IX orbit.)
\end{conjecture}

The Mixmaster stochasticity conjecture is based on a rather
suggestive simple idea: Type~IX evolution is like trying to
follow a path of an (infinite) network of paths while the
ground is shaking randomly, and where the shaking subsides with
time but never stops. A type~IX orbit tries to follow a
sequence of transitions on the Mixmaster attractor while random
errors cause the orbit to lose track. Due to the errors, the
orbit is incapable of following one particular sequence
forever; after a finite time,
the type~IX orbit has deviated too much 
and it leaves the vicinity of the sequence.
Although, temporarily, the orbit is contained in a
neighborhood of a different sequence, it is bound to
lose track of that sequence as well eventually.
Accordingly the type~IX orbit is thrown around in the space of
Mixmaster sequence with the effect that the type~IX orbit
inherits the stochastic properties of generic Mixmaster
sequences.

In the following we paint a heuristic picture that makes
this idea a little more concrete. 
Consider a type~IX orbit (approximate
sequence of transitions) and the associated approximate
Mixmaster sequence
$(\checkP_l)_{l\in\mathbb{N}}$. 
Let $(\mathrm{P}^{(0)}_l)_{l\in\mathbb{N}}$ be the exact
Mixmaster sequence with $\mathrm{P}_0^{(0)} = \checkP_0$ and
consider the Taub-adapted neighborhood of this sequence
associated with some prescribed value $\epsilon > 0$. Finite
Mixmaster shadowing entails that there exists a finite piece
$(\checkP_l)_{l=0,1,\ldots, L_1-1}$ of the approximate sequence
that is contained in the prescribed Taub-adapted neighborhood
of the exact sequence $(\mathrm{P}^{(0)}_l)_{l\in\mathbb{N}}$.
However, at $l = L_1$, the approximate Mixmaster sequence
$(\checkP_l)_{l\in\mathbb{N}}$ leaves the prescribed tolerance
interval due to the accumulation of errors. Hence, at $l = L_1$
we reset the system and consider the exact Mixmaster sequence
$(\mathrm{P}^{(1)}_l)_{l \geq L_1}$ with initial data
$\mathrm{P}^{(1)}_{L_1} = \check{\mathrm{P}}_{L_1}$. The
approximate Mixmaster sequence $(\checkP_l)_{l \geq L_1}$ is
contained in the Taub-adapted neighborhood of the exact
sequence $(\mathrm{P}^{(1)}_l)_{l \geq L_1}$ up to $l = L_2
-1$. At $l = L_2$ another readjustment becomes necessary.
Iterating this procedure and concatenating the finite pieces
$(\mathrm{P}^{(i)}_l)_{l=L_i,\ldots,L_{i+1}-1}$ we are able to
construct a sequence $(\bar{\mathrm{P}}_l)_{l\in\mathbb{N}}$
with the property that the approximate sequence
$(\checkP_l)_{l\in\mathbb{N}}$ is contained within the
Taub-adapted neighborhood (associated with $\epsilon$) of
$(\bar{\mathrm{P}}_l)_{l\in\mathbb{N}}$ for all $l
\in\mathbb{N}$. The sequence
$(\bar{\mathrm{P}}_l)_{l\in\mathbb{N}}$ is a piecewise exact
Mixmaster sequence; it is exact in intervals $[L_i, L_{i+1})$.
The length $\norDelta L_i = L_{i+1} - L_i$ of these intervals
grows beyond all bounds as $i\rightarrow \infty$, because
shadowing takes place with an increasing degree of accuracy.
The error (of the order $\epsilon$) between the approximate
sequence $(\checkP_l)_{l\in\mathbb{N}}$ and the piecewise exact
sequence $(\bar{\mathrm{P}}_l)_{l\in \mathbb{N}}$ results from
the accumulation of numerous small errors. This obliterates the
deterministic origin of the problem  and generates a
`randomness' that leads to stochastic properties. Accordingly,
we expect that the exact sequences $(\mathrm{P}^{(i)}_l)_{l
\geq L_i}$ from which $(\bar{\mathrm{P}}_l)_{l\in\mathbb{N}}$
is built constitute a random sample of Mixmaster sequences and
thus truly reflect the stochastic properties of the Mixmaster
map. As a consequence, although the sequence
$(\bar{\mathrm{P}}_l)_{l\in \mathbb{N}}$ is only a piecewise
exact Mixmaster sequence, 
it possesses the same stochastic
properties as a generic Mixmaster sequence.
Extrapolating this line of reasoning we are able to
complete the argument and find that the approximate sequence
$(\check{\mathrm{P}}_l)_{l\in\mathbb{N}}$ itself
reflects the stochastic properties of the Mixmaster map.
To emphasize this aspect of stochasticity of $(\check{\mathrm{P}}_l)_{l\in\mathbb{N}}$
we use the term `randomized approximate sequence'.
Some comments are in order.

In our discussion we have assumed implicitly that the
approximate sequence $(\check{\mathrm{P}}_l)_{l\in\mathbb{N}}$
we consider can shadow any exact Mixmaster sequence for a
finite number of transitions only. This is not necessarily the
case. A type~IX orbit whose $\alpha$-limit set is one of the
heteroclinic cycles (if such an orbit exists!) is an obvious
counterexample: For every $\epsilon > 0$ there exists $L
\in\mathbb{N}$ such that $\check{\mathrm{P}}_l$ is contained in
the Taub-adapted $\epsilon$-neighborhood of the Mixmaster
sequence associated with the cycle. However, we expect that
this type of `infinite shadowing' holds at most for orbits of a
set of measure zero.

A more serious limitation of the intuitive picture that we have
sketched is illustrated by the following related example.
Consider a type~IX orbit whose $\alpha$-limit set is the
heteroclinic cycle depicted in Figure~\ref{period1a} (where we
note again that the existence of such an orbit is not proven).
For the associated approximate Kasner sequence
$(\check{u}_l)_{l\in\mathbb{N}}$ we have $\check{u}_l
\rightarrow (1+\sqrt{5})/2$ as $l\rightarrow \infty$. Consider
the piecewise exact Kasner sequence
$(\bar{u}_l)_{l\in\mathbb{N}}$ that is associated with the
piecewise exact Mixmaster sequence. Each piece
$(u_l^{(i)})_{l=L_i,\ldots,L_{i+1}-1}$ is generated from a
value $u_{L_i}^{(i)} = [k_0;k_1,k_2,\ldots] =
[1;1,1,1,1,\ldots,1,k^{(i)}_n,k^{(i)}_{n+1},\ldots]$; we have
$n\rightarrow \infty$ as $i\rightarrow \infty$. It is evident
that these pieces do \textit{not} form a random sample of
Kasner sequences. At best one could conjecture (and it is
probably safe to do so) that the collection of the remainders
$[k^{(i)}_n;k^{(i)}_{n+1},\ldots]$ is such a random sample.
Even so, the stochastic aspect does not carry over to the
approximate sequence $(\check{u}_l)_{l\in\mathbb{N}}$, since
the approximate sequence leaves the neighborhood of each
sequence $(u_l^{(i)})_{l=L_i,\ldots}$ before that sequence has
entered its stochastic regime (which is characterized by the
remainder $[k^{(i)}_n;k^{(i)}_{n+1},\ldots]$). Again we invoke
`genericity' to save the day: We conjecture that, for almost
all approximate sequences, the sequences $(u^{(i)}_l)_{l \geq
L_i}$ are a true random sample of Kasner sequences and that the
associated stochastic properties indeed carry over to the
approximate sequence $(\check{u}_l)_{l\in\mathbb{N}}$ itself.

The piecewise exact sequence
$(\bar{\mathrm{P}}_l)_{l\in\mathbb{N}}$ consists of pieces of
length $\norDelta L_i = L_{i+1} - L_i$, where $L_i$ grows beyond
all bounds as $i\rightarrow \infty$, because shadowing takes
place with an increasing degree of accuracy. However, this does
not necessarily imply that $\norDelta L_i \rightarrow \infty$ as
$i\rightarrow \infty$. The latter property is directly
connected with the \textit{Kasner map convergence conjecture}.

\begin{conjecture}[Kasner map convergence]
Almost every generic type~IX orbit 
is associated with an approximate Kasner sequence
$(\check{u}_l)_{l\in\mathbb{N}}$ such that
$\norDelta\check{u}_l\rightarrow 0$ as $l\rightarrow \infty$,
where $\norDelta\check{u}_l$ describes the error between
$\check{u}_{l+1}$ and the value of the Kasner parameter
generated from $\check{u}_l$ by the exact Kasner
map~\eqref{Kasnermap}.
\end{conjecture}

The relevance of the Kasner map for the asymptotic evolution
of type~IX solutions rests on the validity of
the Kasner map convergence conjecture.
Let us thus give a more detailed discussion and
present the line of arguments that leads to
this conjecture.

An orbit $X(\tau)$ (with segments $X(\tau_l) \rightarrow X(\tau_{l+1})$, $l\in\mathbb{N}$)
generates a sequence of check points
$\cdots \mapsto \check{\mathrm{P}}_l \mapsto \check{\mathrm{P}}_{l+1} \mapsto \cdots$,
which in turn yields a map $\cdots \mapsto \check{u}_l\mapsto \check{u}_{l+1}\mapsto \cdots$.
The parameter $\check{u}_{l+1}$ (associated with
$\check{P}_{l+1}$) is generated from $\check{u}_l$ (associated
with $\check{P}_{l}$) by the Kasner map plus an error
$\norDelta\check{u}_l$, see Figure~\ref{checkmap}. The
magnitude of the error $\norDelta \check{u}_l$ depends on the
initial data of the segment, i.e., on $X(\tau_l)$;
equivalently, we may view $\norDelta\check{u}_l$ as a function
depending on (i) the position of $\checkP_l$ on
$\mathrm{K}^\ocircle$, and (ii) the
vector 
$X(\tau_l) - \checkP_l$,
which is orthogonal to
$\mathrm{K}^\ocircle$ at $\checkP_l$.

To obtain an estimate for $\norDelta \check{u}_l$ we introduce the
\textit{order of magnitude} of the error, which we denote by
$\barDelta\check{u}_l$. We define $\barDelta\check{u}_l$
to be the average of $|\norDelta \check{u}_l|$ over all vectors
$X(\tau_l) - \checkP_l$
of equal length
that are orthogonal to $\mathrm{K}^\ocircle$; alternatively, we
use the somewhat `safer' definition of $\barDelta\check{u}_l$
as the maximum of $|\norDelta \check{u}_l|$. By design, the
order of magnitude $\barDelta\check{u}_l$ of the error is a
function of two variables: (i) the position of $\checkP_l$ on
$\mathrm{K}^\ocircle$, which is invariantly represented by
$\check{u}_l$, and, instead of $X(\tau_l) - \checkP_l$
itself, (ii) the (orthogonal) distance of $X(\tau_l)$ from
$\checkP_l$ (or, equivalently, from $\mathrm{K}^\ocircle$),
i.e.,
$\|X(\tau_l) - \checkP_l\|$ ($= \| X(\tau_l) - \mathrm{K}^\ocircle \|$),
which we denote by
$\check{\delta}_l$, see Figure~\ref{checkmap};
in brief, $\barDelta\check{u}_l = (\barDelta\check{u})(\check{u}_l, \check{\delta}_l)$.
In the following we investigate the
behavior of the function $(\barDelta\check{u})(\check{u},\check{\delta})$
under variations of the two arguments.

(i) $(\barDelta\check{u})(\check{u},\cdot)$.
Keep $\check{u}$ fixed
(and assume that $\check{u}$ lies
in the interval $\check{u}\in (1,\infty)$ so that its image under the Kasner map
is finite). Then $\barDelta \check{u}$ is a function of the
distance 
$\check{\delta}$ such that $\barDelta \check{u} \rightarrow 0$
as $\check{\delta} \rightarrow 0$ (which is a simple
consequence of the regularity of the dynamical system and continuous
dependence on initial data).
It is reasonable to conjecture that the relation is in fact monotone.

\begin{figure}[ht]
\psfrag{a}[cc][cc][0.7][0]{$\check{u}=1$}
\psfrag{u}[cc][cc]{$\check{u}$}
\psfrag{du}[cc][cc]{$\barDelta\check{u}$}
\psfrag{d}[cc][cc][1][0]{$\check{\delta}$}
\centering{\includegraphics[width=0.5\textwidth]{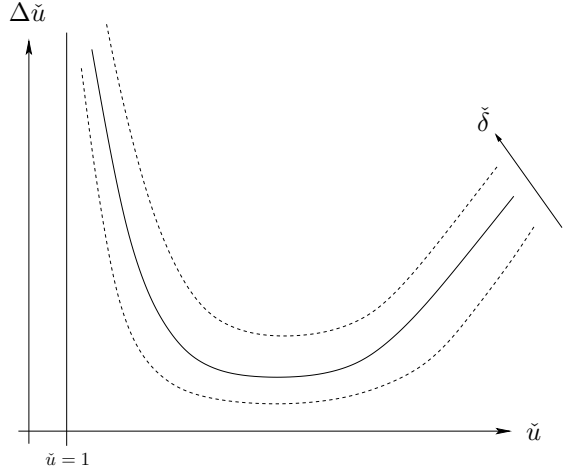}}
\caption{We consider a segment of an orbit $X(\tau)$ with initial check point
$\checkP$ associated with the Kasner value $\check{u}$.
(In accordance with the standard convention for the Mixmaster/Kasner map,
the terms `initial' and `final' refer to a past-directed time.)
The quantity $\check{\delta}$ denotes
the initial distance of the orbit from the Kasner circle, which coincides
with its distance from $\checkP$.
The final check point is not generated from $\checkP$ via the
Mixmaster map, but with an error 
represented by $\norDelta\check{u}$.
The figure gives a schematic depiction of the order of
magnitude $\barDelta\check{u}$ of the error
in dependence on $\check{u}$ and $\check{\delta}$.
Each curve represents $(\barDelta\check{u})(\cdot,\check{\delta})$,
i.e., $\barDelta\check{u}$ as a function of $\check{u}$
for a constant value of the initial distance $\check{\delta}$;
the lower curve is associated with a smaller value of $\check{\delta}$,
the top curve with a larger value.}
\label{uerror}
\end{figure}

(ii) $(\barDelta\check{u})(\cdot,\check{\delta})$.
Keep the distance $\check{\delta}$ fixed,
so that $\barDelta \check{u}$ is a function of
$\check{u}$.
The fundamental observation is that $\barDelta\check{u}$ becomes unbounded
as (a) $\check{u}\rightarrow \infty$, and (b) $\check{u}\rightarrow 1$.
Case (a) is due to the intricacies of the flow in the vicinity
of the non-transversally hyperbolic Taub points (where $u = \infty$);
recall that $\check{u}^{-1}$ measures the angular distance
of the check point from the nearest Taub point.
For (b) we consider the orbit $\mathrm{Q}_\alpha \rightarrow
\mathrm{T}_\alpha$ (for some $\alpha$) which corresponds to $u
= 1 \mapsto u =\infty$. Suppose that $\checkP_l$ coincides with
$\mathrm{Q}_\alpha$ (i.e., $\check{u}_l = 1$).
In general, $\checkP_{l+1}$ will not coincide with
$\mathrm{T}_\alpha$ (independently of the choice
of $\check{\delta} = \check{\delta}_l$);
therefore $\barDelta \check{u}_l = (\barDelta\check{u})(\check{u}_l, \check{\delta}_l) = \infty$.
If $\checkP_l$ is very close to
$\mathrm{Q}_\alpha$, which means that $\check{u}_l$ is close to
$1$, a small deviation of $X(\tau)$ from the type~II transition
emanating from $\checkP_l$ will still be small at the end point
$X(\tau_{l+1})$; however, in the vicinity of the Taub point
$\mathrm{T}_\alpha$ even a small deviation can translate to a
large error $\norDelta \check{u}_{l}$ between $\check{u}_{l+1}$
and $(\check{u}_l-1)^{-1}$, which in turn results in
the asserted blow-up of the order of magnitude of the error.
%
The qualitative properties of $\barDelta\check{u}$ as a function of
$\check{\delta}$ and 
$\check{u}$ are depicted in Figure~\ref{uerror}.

Consider a generic type~IX orbit $X(\tau)$ and its
representation as an approximate sequence of transitions. By
Theorem~\ref{rinthm} each orbit converges to the Mixmaster
attractor. As a consequence the distance $\check{\delta}_l =
\|X(\tau_l) - \mathrm{K}^\ocircle \|$ ($ = \| X(\tau_l) -
\checkP_l \|$) converges to zero as $l\rightarrow \infty$.
Therefore, as $\check{\delta}_l \rightarrow 0$ ($l\rightarrow\infty$),
the sequence of functions $(\barDelta\check{u})(\cdot,\check{\delta}_l)$
of Figure~\ref{uerror} converges to zero. However, this convergence
is merely pointwise and not uniform;
this behavior is crucial to understand the behavior
of $\norDelta\check{u}_l$ as $l\rightarrow \infty$.

If a type~IX orbit $X(\tau)$ converges to one of the
heteroclinic cycles on
the Mixmaster attractor, 
then there exists $\varepsilon$ such that $\check{u}_l \in
(1+\varepsilon,\varepsilon^{-1})$ for all sufficiently large
$l$. On this interval the function $\barDelta\check{u}$ of
Figure~\ref{uerror} converges to zero uniformly. We therefore
obtain that $\barDelta\check{u}_l = (\barDelta\check{u})(\check{u}_l,\check{\delta}_l)
\rightarrow 0$ and thus
$\norDelta \check{u}_l \rightarrow 0$ as $l\rightarrow \infty$ for
this type~IX orbit. However, the Mixmaster attractor conjecture
suggests that almost every type~IX orbit $X(\tau)$ has an
associated approximate Kasner sequence
$(\check{u}_l)_{l\in\mathbb{N}}$ that is unbounded, hence
the general case is not so clear. Since
$(\check{u}_l)_{l\in\mathbb{N}}$ enters the intervals
$[1,1+\varepsilon)$ and $(\varepsilon^{-1},\infty)$ for any
$\varepsilon$, $\barDelta\check{u}_l = (\barDelta\check{u})(\check{u}_l,\check{\delta}_l)$
(and thus $\norDelta\check{u}_l$) need not necessarily converge to zero.
Let us elaborate.

Let $\kappa > 0$ be fixed.
By the implicit function theorem,
the inequality $(\barDelta\check{u})(\check{u},\check{\delta}) \geq \kappa$
is satisfied if and only if
$\check{u} \leq 1 + f_\kappa(\check{\delta})$
or $\check{u} \geq g_\kappa(\check{\delta})$
for some functions $f_\kappa$, $g_\kappa$, which satisfy
$f_\kappa(\check{\delta}) \rightarrow 0$
and $g_\kappa(\check{\delta}) \rightarrow \infty$ as $\check{\delta}\rightarrow 0$.
Accordingly,
for a given type~IX orbit (and its associated approximated
Kasner sequence $\check{u}_l$), we obtain
\begin{equation}\label{orderhazard}
\barDelta\check{u}_l = (\barDelta\check{u})(\check{u}_l,\check{\delta}_l) \geq \kappa
\quad\Leftrightarrow\quad
\check{u}_l \leq 1 + f_\kappa(\check{\delta}_l)
\end{equation}
or $\check{u}_l \geq g_\kappa(\check{\delta}_l)$.
We call the union of the two
intervals $\big(1, 1 +  f_\kappa(\check{\delta}_l)\big)$ and $\big(g_\kappa(\check{\delta}_l),\infty\big)$
the `hazard zone' (associated with epoch number $l$);
obviously, the hazard zone is decreasing with $l$.
Hence, the order of magnitude of the error, $\barDelta\check{u}_l$,
is larger than $\kappa$
at epoch number $l$ if and only if the approximate Kasner
parameter $\check{u}_l$ falls into the hazard zone (associated with $l$).

In connection with the Kasner map convergence conjecture the question is
how often $(\check{u}_l)_{l\in\mathbb{N}}$ enters the hazard zone,
i.e., how often~\eqref{orderhazard} occurs: finitely many
times or infinitely many times?

Suppose that the Mixmaster stochasticity conjecture is correct.
Then almost every sequence $(\check{u}_l)_{l\in\mathbb{N}}$
admits a probabilistic interpretation in terms of the
probability distribution~\eqref{Khinchin}. Accordingly, we
expect the question to turn into an example of a `0--1 law':
There exists two alternatives: (i) Type~IX orbits (i.e.,
sequences $(\check{u}_l)_{l\in\mathbb{N}}$) that
satisfy~\eqref{orderhazard} infinitely many times are generic
(i.e., of measure~$1$ in the state space); type~IX orbits that
do not are non-generic (i.e., of measure~$0$); (ii) Type~IX
orbits that satisfy~\eqref{orderhazard} infinitely many times
are non-generic; orbits that do not are generic. (In both
cases, the two sets, the `generic set' and the `non-generic
set', are probably non-empty.) Which of these alternatives is
actually realized depends on the rate of decay of
$(\check{\delta}_l)_{l\in\mathbb{N}}$ and the decay and growth
of $f_\kappa$ and $g_\kappa$, respectively, as
$\check{\delta}\rightarrow 0$. Case (i) is associated with
small (subcritical) rates of decay; case (ii) with large
(overcritical) decay. We conjecture that case (ii) applies,
i.e.,~\eqref{orderhazard} is satisfied only finitely many times
for generic type~IX orbits. The reason for this conjecture is
the fast rate of convergence of the orbit to the Mixmaster
attractor which follows from the decay of the functions $N_1
N_2 N_3$, cf.~\eqref{Deltadef}, and~$N_1 N_2 + N_2 N_3 + N_3
N_1$, cf.~\cite{rin01}; this leads to the expectation that the
r.h.\ side of~\eqref{orderhazard} represents a rapidly
decaying function.%
\footnote{Additional support for the conjecture comes from toy models that
  reflect the instability of the Kasner map and its consequences.
  We will come back to this issue in future work.}
To conclude our line of arguments in favor of the Kasner map convergence
conjecture, we note that
the actual error $\norDelta\check{u}_l$ can be estimated by
$\barDelta\check{u}_l$.
Therefore, if $\barDelta\check{u}_l < \kappa$ for all
$l$ except for a finite set, then also $\norDelta\check{u}_l < \kappa$
for all $l$ except for a finite set.
Since $\kappa$ is an arbitrary positive number, the statement
of the Kasner map convergence conjecture ensues.

We round off this section with some further remarks on the conjectures.
First, we note that
one might be led to suppose that there could in fact exist type~IX
solutions for which the statement of the Kasner map convergence
conjecture is violated,
i.e., $\norDelta\check{u}_l \not\rightarrow 0$ as $l\rightarrow
\infty$.
In any case, the class of
these special solutions is at most of measure zero.
Second, we note that the statement of the Kasner map convergence
conjecture, i.e., $\norDelta\check{u}_l \rightarrow 0$
($l\rightarrow \infty$), and the statement
$\norDelta L_i \rightarrow \infty$ ($i\rightarrow \infty$),
cf.~the discussion on piecewise exact sequences,
are expressions of one and the same stochastic property.
This emphasizes that the two `stochastic' conjectures
tightly intertwine.
Third, we remark that
it is conceivable that `almost every' in the two conjectures
might be in fact `every'. In that case,
the non-generic examples of
Kasner/Mixmaster sequences, see Section~\ref{maps}, 
would not have any counterparts among the type~IX solutions.
Of particular interest in this context would be to
have an answer to question (ii) of the second subsection of
Section~\ref{beliefs}. If there do not exist type~IX orbits
whose $\alpha$-limit set is one of the heteroclinic cycles,
this would be a strong indication in favor of `every' and
against `almost every'. If there exist type~IX orbits that
converge to a cycle, `almost every' is the best one can aim for
in the Mixmaster stochasticity conjecture.

Finally, let us briefly comment on claims of stochastic and
chaotic properties of Bianchi type~IX asymptotic dynamics.
Chaotic aspects of the Kasner map and related maps have been
studied under various aspects~\cite{khaetal85, bar82, rugh90,
corlev97, motlet01}. However, the relevance of these maps for
type~IX asymptotics rests on the two conjectures in this
section. If the conjectures turn out to be wrong (e.g., if
there exists a generic set of solutions that converge to
heteroclinic cycles), then none of the results on the Kasner
map carry over to full type~IX dynamics. Numerical
investigations, see~\cite{ber02} and references therein,
reflect Theorem~\ref{rinthm} and finite shadowing, but it is
implausible that numerics can possibly shed light on the
actual asymptotic limit of type~IX solutions. Numerical errors
are unavoidable and random in nature; these errors generate
precisely the type of stochasticity the simulation is looking
for. Accordingly, numerical studies will necessarily find
type~IX solutions to exhibit the same stochastic behavior as
generic Kasner sequences and are thus not suited to explore the
validity of the conjectures.

\section{Billiards and billiard beliefs}
\label{billiard}

In this section we briefly review the Hamiltonian billiard approach to
type~IX dynamics in the spirit of~\cite{chi72,dametal03}
and make contact with the dynamical systems approach.
The metric~\eqref{threemetric} can be written as
\begin{equation}
{}^4\mathbf{g}  = -(\det g) \nt^2 d x^0\otimes d x^0 +
e^{-2 b^1} \:\hat{\bom}^1\otimes \hat{\bom}^1 +
e^{-2 b^2}\:\hat{\bom}^2\otimes \hat{\bom}^2 +
e^{-2 b^3}\:\hat{\bom}^3\otimes \hat{\bom}^3\:,
\end{equation}
where $x^0$ is an arbitrary time variable that is related to
proper time $t$ by a densitized lapse function $\nt$ according to
$dt = \sqrt{\det g} \,\nt\,dx^0$;
evidently, 
$\det g = \exp[-2(b^1 + b^2 + b^3)]$.
The Hamiltonian for the orthogonal Bianchi type IX
perfect fluid models is given by
\begin{equation}\label{Hamilto}
\mathcal{H} = \nt \left( \textfrac{1}{4}
\sum_{{\alpha},\beta}\cg^{{\alpha} \beta} \pi_{\alpha} \pi_{\beta} -
{}^{3}\!R \:\det g  + 2 \rho \:\det g \right) = 0\:,
\end{equation}
where ${}^3\!R$ is three-curvature and
$\rho = \rho_0 (\det g)^{-(1+w)/2}$, cf.~the remark
following~\eqref{Omegaeq} (where we recall that $d \log (\det g) = 6 d\tau$).
$\cg^{{\alpha} \beta}$ is the inverse of the so-called
minisuperspace metric $\cg_{{\alpha} \beta}$,
\begin{equation}
\sum_{{\alpha},\beta} \cg_{{\alpha} \beta} v^{\alpha} w^{\beta}  :=
-\sum_{\gamma\neq \delta} v^{\gamma} w^{\delta} = \sum_{\alpha}
v^{\alpha} w^{\alpha} - \Big( \sum_{\alpha} v^{\alpha} \Big) \Big(
\sum_{\beta} w^{\beta} \Big)\:,
\end{equation}
i.e., $\cg_{\alpha\beta}$ is a $2+1$-dimensional Lorentzian
metric. The gravitational potential $U_G = -^{3}\!R\,\det g$ is
given by
\begin{equation} U_G = \sfrac{1}{2} \left( e^{-4b^1} +
e^{-4b^2} + e^{-4b^3} -2 e^{-2(b^1 + b^2)} - 2 e^{-2(b^2 + b^3)} - 2
e^{-2(b^3 + b^1)} \right)\:;
\end{equation}
where we have set the structure constants $\hat{n}^\alpha$ to
one; the potential for the fluid is given by $U_F = 2
\rho\,\det g = 2\rho_0\exp[-(1-w)(b^1 + b^2 + b^3)]$. For
further details see, e.g.,~\cite{jan01} or~\cite[Chapter
10]{waiell97}; note that $b^\alpha=-\beta^\alpha$, which is
used in these references; see also~\cite{heietal07} and
\cite{dametal03}.

As argued in~\cite{mis69b,dametal03}, $b^{\alpha}$ is expected
to be timelike in the asymptotic regime, i.e., $\sum_{\alpha}
\cg_{\alpha\beta}\, b^{\alpha}b^{\beta} <0$. Assuming that
$b^{\alpha}$ is timelike allows to introduce new metric
variables instead of $\{b^1, b^2,b^3\}$. Defining $\bar{\rho}^2
= -\sum_{\alpha} \cg_{\alpha\beta}\, b^{\alpha} b^{\beta}$ and
orthogonal angular metric variables, collectively denoted by
$\gamma$, leads to
\begin{equation}
\sum_{{\alpha},\beta} \cg_{\alpha\beta}\, d
b^{\alpha} d b^{\beta} = -d \bar{\rho}^2 + \bar{\rho}^2 d\Omega_h^2\:,
\end{equation}
where $d\Omega_h^2$ is the standard metric on hyperbolic space.
Making a further change of variables,
\begin{equation}\label{lambdarho}
\lambda = \log \bar{\rho} = \textfrac{1}{2} \log
\Big(-\sum_{{\alpha},\beta} \cg_{\alpha\beta}\, b^{\alpha} b^{\beta}
\Big)\:,
\end{equation}
yields
\begin{equation}
\sum_{{\alpha},\beta}\cg^{\alpha\beta}\, \pi_{\alpha} \pi_{\beta} =
-\pi_{\bar{\rho}}^2 + (\bar{\rho})^{-2} \pi_\gamma^2 = (\bar{\rho})^{-2} \left[
-\pi_\lambda^2 + \pi_\gamma^2 \right]\:.
\end{equation}

Choosing the lapse according to ${\tilde N}=\bar{\rho}^2$ leads
to a Hamiltonian of the form
\begin{equation}\label{Hami}
{\cal H} = \textfrac{1}{4} \left[ -\pi_\lambda^2 + \pi_\gamma^2
\right] + \bar{\rho}^2 \sum_A c_A\,\exp(-2 \bar{\rho}\, w_A(\gamma))\:,
\end{equation}
where $w_A(\gamma)$ denotes
certain linear forms of the variables $\gamma^{\alpha}$, i.e.,
$w_A(\gamma) = \sum_{\beta} w_{A\,\beta} \,\gamma^{\beta}$;
see~\cite{dametal03} for details.

The essential point is that one expects that
$\bar{\rho}\rightarrow \infty$ towards the singularity and
hence that each term $\bar{\rho}^2 \exp[- 2\bar{\rho}
w_A(\gamma)]$ becomes an infinitely high sharp wall described
by an infinite step function $\Theta_\infty(x)$ that vanishes
for $x<0$ and is infinite for $x\geq 0$. Accordingly, only
`dominant' terms in the potential are assumed to be of
importance for the generic asymptotic dynamics, while
`subdominant' terms, i.e., terms whose exponential functions can
be obtained by multiplying dominant wall terms, are neglected.
In the present case there are three dominant terms in $U_G$
(which is the minimal set of terms required to 
define the billiard table),
the three exponentials $\exp(-4b^\alpha)$. Dropping the
subdominant terms in the limit $\bar{\rho} \rightarrow \infty$
leads to an asymptotic Hamiltonian of the form
\begin{equation}\label{limiHami}
{\cal H}_{\infty} =  \textfrac{1}{4} \left[ -\pi_\lambda^2 +
\pi_\gamma^2 \right] + \sum_{A=1}^3 \Theta_\infty(-2 w_A(\gamma))\:,
\end{equation}
where only the three dominant terms appear
in the sum.
The correspondence between the dynamical systems picture and
the Hamiltonian picture is easily obtained by noting that the
Hamiltonian constraint~\eqref{Hamilto} is proportional to the
Gauss constraint~\eqref{gauss}. The dominant terms correspond
to the terms $N_\alpha^2$, $\alpha =1,2,3$, in~\eqref{gauss};
the subdominant terms are collected in~$\Delta_{\mathrm{II}}$.

\begin{figure}[ht]
\psfrag{a}[cc][cc]{$\gamma_1$} \psfrag{b}[cc][cc]{$\gamma_3$}
\psfrag{c}[cc][cc]{$\gamma_2$} \ \centering{
  \includegraphics[height=0.45\textwidth]{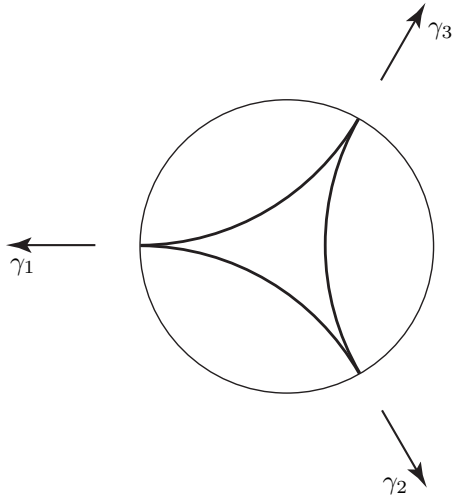}}
\caption{The Bianchi type IX billiard: The disc represents
hyperbolic space. The asymptotic description of a solution is given by
geodesic billiard motion inside the triangle, which acts as
a stationary infinite potential wall, yielding a `configuration
space picture' of the asymptotic dynamics.}\label{billiardIX}
\end{figure}

The non-trivial dynamics described by~\eqref{limiHami} resides
in the variables $\gamma$, i.e., in the hyperbolic space. It
can be described asymptotically as geodesic motion in
hyperbolic space constrained by the existence of sharp
reflective walls, i.e., the asymptotic dynamics is determined
by the type IX `billiard' given in Figure~\ref{billiardIX}.
Based on the heuristic considerations the limiting
Hamiltonian~\eqref{limiHami} is believed to describe generic
asymptotic dynamics.

The Hamiltonian picture (as represented by the billiard) and
the dynamical systems picture (as represented by the Mixmaster
attractor) are `dual' to each other. In the Hamiltonian picture
the asymptotic dynamics is described as the evolution of a
point in the billiard. Straight lines (geodesics) in hyperbolic
space correspond to Kasner states. Wall bounces correspond to
Bianchi type~II solutions; the bounces change the Kasner states
according to the Kasner map. Since the billiard picture
emphasizes the dynamics of the configuration space variables,
one may say that the Hamiltonian billiard approach yields a
`configuration space' representations of the essential
asymptotic dynamics.

In the dynamical systems state space picture the motion in the
Hamiltonian billiard picture becomes a `wall' of fixed
points---the Kasner circle $\mathrm{K}^{\ocircle}$. The walls
in the Hamiltonian billiard are translated to motion in the
dynamical systems picture---Bianchi type~II heteroclinic
orbits; these type~II transitions yield exactly the same rule
for changing Kasner states as the wall bounces: the Kasner map.
Since the variables $\Sigma_\alpha$ are proportional to
$\pi_\alpha$, it is natural to refer to the projected dynamical
systems picture as a `momentum space' representation of the
asymptotic dynamics.

Summing up, `walls' and `straight line motion' switch places
between the Hamiltonian formulation and dynamical systems
description of asymptotic dynamics, and the two pictures give
equivalent complementary asymptotic pictures.

The above heuristic derivation of the limiting Hamiltonian
rests on two basic assumptions: It is assumed that $b^{\alpha}$
is timelike in the asymptotic regime and that one can drop the
subdominant terms. These assumptions correspond to assuming
that $\Omega$ and $\Delta_{\mathrm{II}}$ can be set to zero
asymptotically, i.e., the procedure precisely assumes
Theorem~\ref{rinthm}. (The Hamiltonian analysis
in~\cite[Chapter 2]{monetal07} uses the same assumptions, and
hence the present discussion is of direct relevance for that
work as well.) That these assumptions are highly non-trivial is
indicated by the difficulties that the proof of
Theorem~\ref{rinthm} has presented; cf.~\cite{rin01,
heiuggproof}. An alarming example is Bianchi type~VIII; in this
case the heuristic procedure in this respect leads to exactly
the same asymptotic results, but so far there exist no proof
that $\Omega$ and $\Delta_{\mathrm{II}}$ tend to zero towards
the singularity for generic solutions. We elaborate on the
differences between type~VIII and type~IX
in~\cite{heiuggproof}.
Moreover, like in the state space picture there exists no proof
that all of the possible billiard trajectories are of relevance
for the asymptotic dynamics of type~IX solutions. For example,
there exists a correspondence between periodic orbits in the
dynamical systems picture and the Hamiltonian billiard picture,
and it is not excluded a priori that solutions are forced to
one, or several, of
these. 
We emphasize again that all proposed measures of chaos that
take billiards as the starting point, see~\cite{monetal07} and
references therein, rely the conjectured connection between the
Mixmaster map and asymptotic type~IX dynamics.

The above discussion shows that there are non-trivial assumptions
and subtle phenomena that are being glossed over in heuristic
billiard `derivations.' Nevertheless, we do believe that the
billiard procedure elegantly uncovers the main generic asymptotic
features, and it might be fruitful to attempt to combine the
dynamical systems and Hamiltonian picture in order to prove the
Mixmaster conjectures.
A tantalizing hint is that in the billiard picture
$\pi_\lambda$ becomes an asymptotic constant of the motion.
Rewriting  this dimensional constant of the motion in
terms of the dynamical systems variables yields
\begin{equation}
\pi_\lambda = 2 H \sqrt{\det g} \left[ -(\tau - \tau_0) +
\textfrac{1}{12} \log \big( N_1^{\Sigma_1}  N_2^{\Sigma_2} N_3^{\Sigma_3}\big) \right]\:,
\end{equation}
where $\log \sqrt{\det g} = 3 (\tau -\tau_0)$ and
$\tau_0$ is a constant.

\section{Concluding remarks}
\label{concl}

The purpose of this paper is two-fold. On the one hand, we
analyze the main known results on the asymptotic dynamics of
Bianchi type~IX vacuum and orthogonal perfect fluid models
towards the initial singularity. The setting for our discussion
is the Hubble-normalized dynamical systems approach, since this
is essentially the set-up which has led to the first rigorous
statements on Bianchi type~IX asymptotics~\cite{rin01}. (We
choose slightly different variables to emphasize the
permutation symmetry that underlies the problem.) The main
result (`Mixmaster fact') is Theorem~\ref{rinthm} which is due
to Ringstr\"om~\cite{rin01}; for an alternative proof
see~\cite{heiuggproof}. This theorem, in conjunction with an
analysis of the Mixmaster attractor, leads to a number of
further rigorous results which we list as
consequences~\cite{heiuggproof}.

On the other hand, we draw a clear line between rigorous
results (`facts') and heuristic considerations (`beliefs'). We
make explicit that the implications of Theorem~\ref{rinthm} and
its consequences are rather limited, in particular, the
rigorous results do not give any information on the details of
the oscillatory nature of Mixmaster asymptotics. The
mathematical methods required to obtain proofs about the actual
asymptotic Mixmaster oscillations are yet to be developed; it
is likely that radically new ideas are needed. This paper
provides the infrastructure that might yield the basis for
further developments. Our framework enables us to transform
vague beliefs to a number of specific conjectures that describe
the expected `complete picture' of Bianchi type~IX asymptotics.
The arguments we give in their favor are based on an in-depth
analysis of the Mixmaster map and its stochastic aspects in
combination with the dynamical systems concept of shadowing.

We conclude with a few pertinent comments. First, as elaborated
in~\cite{heiuggproof} there exists no corresponding theorem to
Theorem~\ref{rinthm} in other oscillatory Bianchi models such
as Bianchi type VI$_{-1/9}$ or VIII; this suggests that the
situation in the general inhomogeneous case is even more
complicated than expected. Furthermore, numerical studies are
incapable of shedding light on the asymptotic limit. This is
mainly due to the accumulation of inevitable random numerical
errors that make it a priori impossible to track a particular
type~IX orbit. Finally, although the Hamiltonian methods are a
formidable heuristic tool, so far this approach has not yielded
any proofs about asymptotics. Nevertheless, it might prove to
be beneficial to explore the possible synergies between
dynamical systems and Hamiltonian methods.

In this paper and
in~\cite{heiuggproof,heietal07} we have encountered remarkable
subtleties as regards the asymptotic dynamics of oscillatory
singularities; this emphasizes the importance of a clear
distinction between facts and beliefs.

\subsection*{Acknowledgments}
We thank Alan Rendall, Hans Ringstr\"om, and in particular 
Lars Andersson for useful discussions. We
gratefully acknowledge the hospitality of the Mittag-Leffler
Institute, where part of this work was completed. CU is
supported by the Swedish Research Council.

\bibliographystyle{plain}

\vfill

\end{document}